\documentclass[referee,a4paper,fleqn,usenatbib]{mnras}
\usepackage[T1]{fontenc}
\usepackage{ae,aecompl}
\usepackage{graphicx}	
\usepackage{amsmath}	
\usepackage{amssymb}	
\usepackage{longtable}
\usepackage[para]{threeparttable}
\usepackage[normalem]{ulem}
\usepackage{lmodern}
\usepackage{color}

\newcommand{\hii}{\mbox{H~{\sc ii}~}}

\title[Variable Stars towards NGC 2282]{Optical Photometric Variable Stars towards the Galactic \hii region NGC 2282}
\author[S. Dutta et al.]{Somnath Dutta$^{1}$\thanks{E-mail: somnath12@boson.bose.res.in (SD)},  Soumen Mondal$^{1}$, Santosh Joshi$^{2}$, Jessy Jose$^{3}$, Ramkrishna Das$^{1}$, \and and Supriyo Ghosh$^{1}$ \\
$^{1}$Satyendra Nath Bose National Centre for Basic Sciences, Kolkata 700 106, India\\
$^{2}$Aryabhatta Research Institute of Observational Sciences, 
Nainital-263002, India \\
$^{3}$Indian Institute of Science Education and Research, Rami Reddy Nagar, Karakambadi Road, Mangalam (P.O.) Tirupati 517507, India}

\begin{document}

\maketitle

\begin{abstract}

We report here CCD $I$-band time-series photometry of a young (2$-$5 Myr) cluster NGC 2282 to identify and understand the variability of pre-main-sequence (PMS) stars. The  $I$-band photometry, down to $\sim$ 20.5 mag, enables us to probe the variability towards the lower mass end ($\sim$ 0.1 M$_\odot$) of the PMS stars. From the light curves of 1627 stars, we identified 62 new photometric variable candidates. Their association with the region was established from  H$\alpha$ emission and infrared (IR) excess. Among 62 variables, 30 young variables exhibit H$\alpha$ emission, near-IR (NIR)/mid-IR (MIR) excess or both, and they are candidate members of the cluster. Out of 62 variables, 41 are periodic variables with the rotation rate ranging from 0.2 to 7 days. The period distribution exhibits a median period at $\sim$ 1-day as in many young clusters (e.g., NGC~2264, ONC, etc.), but it follows a uni-modal distribution unlike others having bimodality  with the slow rotators peaking at $\sim$ 6$-$8 days. To investigate the rotation-disk and variability-disk connection, we derived NIR excess from $\Delta$(I$-$K) and MIR excess from $Spitzer$ [3.6]$-$[4.5] $\mu$m data. No conclusive evidence of slow rotation  with the presence of disks around stars and fast rotation for diskless stars is obtained from our periodic variables. A clear increasing trend of the variability amplitude with the IR excess is found for all variables.

\end{abstract}

\begin{keywords}
Embedded clusters---pre-main sequence -- variable stars.
\end{keywords}

\section{Introduction}

The photometric variability is a ubiquitous characteristic of the young stars \citep[]{1945ApJ...102..168J,1962AdA&A...1...47H}. The variation in observed flux is proposed to be mainly due to the rotational modulation of hot/cool spots on the young star's surface yielding the rotation period ranges from hours to 15 days \citep[]{1994AJ....108.1906H,2001AJ....121.3160C,2005AJ....129..907B}. In a binary system, two components periodically eclipse one another resulting in a change in the apparent brightness of the system. Furthermore, various temporal phenomena such as flare like activity on the corona, circumstellar disk extinction due to the disk asymmetry, variable accretion rates, etc. can lead to aperiodic variability of young stars. Potentially, variability explores the young stars in the field population. Recent studies on PMS variable census allowed  to probe rotation rates in the few Myrs young to several Myrs old clusters those include  the Orion Nebula Cluster \citep[]{1999AJ....117.2941S,2002A&A...396..513H,2010A&A...515A..13R},  Chamaeleon I \citep[]{2003ApJ...594..971J}, IC 348 \citep[]{2004AJ....127.1602C}, $\sigma$ Orionis \citep[]{2004A&A...419..249S,2010ApJS..191..389C},  NGC 2264 \citep[]{2004AJ....127.2228M,2005A&A...430.1005L},  NGC 2362 \citep[]{2008MNRAS.384..675I} and Taurus \citep[]{2009ApJ...695.1648N}. However, variability alone is not sufficient evidence to identify the young members, and sometimes might be false positive.  Consequently, additional constraining tools are required to confirm, such as infrared excess, H$\alpha$ emission, spatial location, position on a color-magnitude diagram (CMD), etc.

\begin{figure*}
\includegraphics[width=16.0 cm,height=14.0cm]{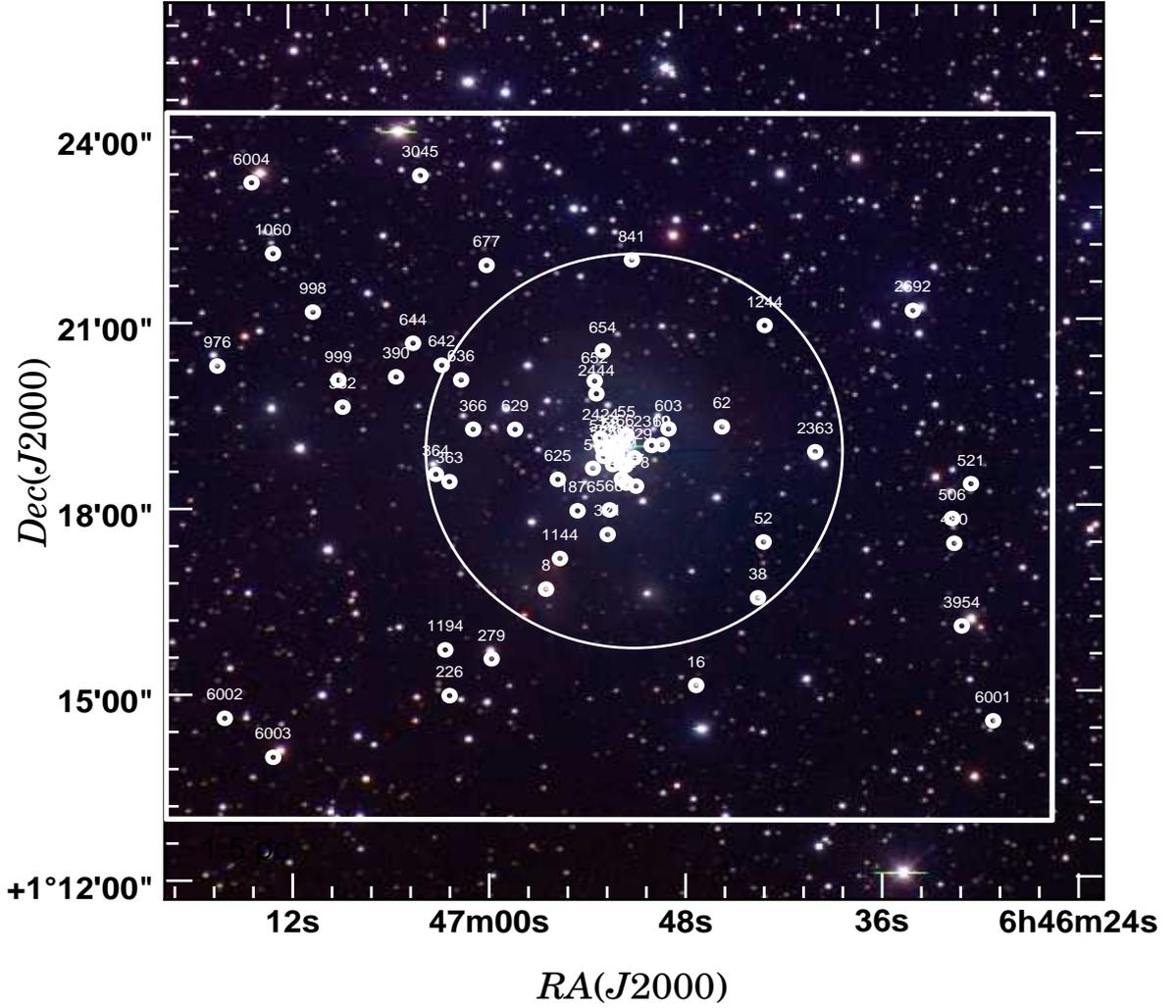}
\caption{Optical color composite image of NGC 2282 (blue: V; green: R; red: I) obtained at 1.3m DFOT (Up is North, left is East). The cluster area is marked with a large white circle. Our combined monitoring observations were done at the white box region. The variable candidates are marked (see text for details).}
  \label{fig:optical_vri}
\end{figure*}

Our principal focus of monitoring program is the discovery of rotation periods and characterizing the variability of PMS stars in young clusters.  Photometric method to study variable stars is a useful technique to pick up membership against field population in such clusters \citep[e.g.,][]{2001Sci...291...93B}. It appears as an effective method to identify Weak-line T Tauri Stars (WTTSs) since they are very challenging to be distinguished from the field stars based on their small infrared access.  The period distribution in some young clusters (a few Myrs old) is bimodal in the period range 1 to 10 days \citep[]{2010ApJS..191..389C,2002A&A...396..513H,2005A&A...430.1005L,2016ApJ...833..122C}, while in others bimodality is not present \citep[]{2004AJ....127.2228M}. The star-disk connection invokes for such period bimodality \citep[]{1991ApJ...370L..39K,1994ApJ...429..781S}. The disk locking mechanism was verified by observations on several young clusters from any correlation between the rotation periods and the disk indicators e.g. $\Delta$(I$-$K), $\Delta$(H$-$K$_s$), [3.6]$-$[8.0] and H$\alpha$  \citep[]{1993AJ....106..372E,1999AJ....117.2941S,2002A&A...396..513H,2004AJ....127.2228M,2005A&A...430.1005L,2006ApJ...646..297R,2007ApJ...671..605C,2010A&A...515A..13R,2010ApJS..191..389C,2012MNRAS.427.1449L}.  However, no  conclusive result was found that those slow rotators are disk-bearing or relatively fast rotators are diskless \citep[]{2007ApJ...671..605C}. Review works in this subject area have explored the rotation-disk connection in this low mass end \citep[]{2007prpl.conf..297H,2007IAUS..243..231B}. Mass-rotation connection reveals a clear mass-dependent morphology, which evolves with the ages with relatively fast rotation seen in the low mass end \citep[][]{2008MNRAS.384..675I,2016ApJ...833..122C}.

 NGC~2282 ($\alpha_{2000}$ = $06^h46^m50.4^s$; $\delta_{2000}$ = $+01^018^m50^s$) is a reflection nebula in the Monoceros constellation, 3 degree away from the Mon OB2 complex and may be associated with it. The cluster parameters of NGC~2282 were first studied by \citet[]{1997AJ....113.1788H} using NIR data. Subsequently, \citet[]{2015MNRAS.454.3597D} (hereafter Paper I) studied more elaborately the PMS member candidates and cluster parameters of NGC~2282 using deep optical $BVIH\alpha$, UKIDSS $JHK$ and $Spitzer$-IRAC data. Using NIR to MIR color-color (CC) diagram and H$\alpha$-emission, the authors identified 152 PMS member candidates, which include 75 Class II and 9 Class I objects. Majority members are concentrated within the radius of $\sim$ 3.15$\arcmin$, and the minimum extinction towards the region is A$_V$ $\sim$ 1.65 mag.  Analyses of the optical and NIR CMDs and disc fraction ($\sim$ 58\%), the age of the cluster was determined to be 2 $-$ 5 Myr. Spectroscopic study of bright stars found that the region contains at least three B-type stars. From optical spectroscopic studies, the authors confirmed the previously known main-illuminating source B2V type star, HD 289120, and identified two more  high mass stars, one Herbig Ae/Be star (B0.5 Ve) and another B5 V star.  From spectroscopy observations, the cluster distance is refined to $\sim$ 1.65 kpc. Thus, NGC 2282 is a suitable object for detailed variability studies as the cluster is relatively populous within a small region, sufficiently nearby and young.

We report here long time-domain photometric studies of NGC~2282 to understand and characterize the variability properties of PMS stars.  Section~2 describes with details of our photometric observations including available archival data for the present work.  Section~3 contains our result and discussion including analysis of the photometric light curves and identifications of new periodic and aperiodic variable young stars. In this section, we also characterize those periodic variables.  Finally, the main results of this work are summarized in section~4.

\begin{table*}

\centering
 \begin{minipage}{140mm}
  \caption{Log of Observations.} 
\label{tab:observation} 
\begin{tabular}{|c|c|c|c|c|}
\hline

 Date of      & Telescope &  I                   & R             & Avg. seeing \\
 Observations &           & Exp.(s)$\times$ N &Exp.(s)$\times$ N &    (arccsec)\\
\hline
\hline
15.10.2013 &2.0m HCT&60$\times$1, 120$\times$30 & 200$\times$1 & 2.5 \\
07.11.2013 &2.0m HCT&120$\times$20, 300$\times$ 1 & 200$\times$1 &2.5 \\
08.11.2013 &2.0m HCT&120$\times$3, 300$\times$ 1 & 300$\times$1 & 2.5 \\
05.01.2014&1.3m DFOT&150$\times$46&250$\times$1 & 1.8 \\
06.01.2014&1.3m DFOT&150$\times$56 & $...$ & 1.9\\
29.03.2014&1.3m DFOT&150$\times$49 &250 $\times$1 & 1.75\\
31.03.2014&1.3m DFOT&150$\times$16, 60$\times$1 & 250$\times$2, 150$\times$1& 1.8\\
05.10.2014&2.0m HCT&90$\times$2 & $...$ &1.5 \\
06.10.2014&2.0m HCT&120$\times$2, 90$\times$4, 60$\times$5 &$...$ & 3.2 \\
29.10.2014&2.0m HCT&120$\times$2, 90$\times$5, 60$\times$5  & $...$& 2.5\\
30.10.2014&2.0m HCT&120$\times$6, 90$\times$8, 60$\times$3  & $...$& 2.5\\
12.11.2014&1.3m DFOT&150$\times$40, 90$\times$27, 60$\times$30 &250$\times$2 & 2.0\\
13.11.2014&1.3m DFOT&150$\times$25, 90$\times$21, 60$\times$20 &250$\times$2 & 2.0\\
14.11.2014&1.3m DFOT&150$\times$35, 90$\times$25, 60$\times$25 &250$\times$2 &2.0 \\
30.11.2014&1.3m DFOT&150$\times$60, 90$\times$41, 60$\times$40  &250$\times$2 &2.0 \\
01.12.2014&1.3m DFOT&180$\times$60, 90$\times$20, 60$\times$20   &250$\times$2 &2.0 \\
10.12.2014&1.3m DFOT& 180$\times$45, 90$\times$10, 60$\times$10  &250$\times$2 &2.0 \\
11.12.2014&1.3m DFOT&180$\times$60, 90$\times$15, 60$\times$15   &250$\times$2 & 2.0\\
12.12.2014&1.3m DFOT&180$\times$50, 90$\times$40, 60$\times$20   &250$\times$2 & 2.0\\
15.01.2015&2.0m HCT&180$\times$3, 120$\times$4, 60$\times$5  & 250$\times$2&2.5 \\
16.01.2015&2.0m HCT&120$\times$3, 60$\times$3  &200$\times$1 &2.5 \\
17.01.2015&2.0m HCT&120$\times$2, 60$\times$2 &200$\times$1 &2.5 \\
05.10.2015&2.0m HCT&150$\times$2, 60$\times$2  &200$\times$1 & 2.5\\
07.10.2015&2.0m HCT&300$\times$2, 60$\times$3 &200$\times$2 & 2.5\\
18.02.2016&2.0m HCT&200$\times$5, 90$\times$6 &200$\times$1, 90$\times$1& 2.5\\

\hline
\hline\end{tabular}
\end{minipage}
\end{table*}

\section[]{Data Sets Used}
\label{data_sets}

The $RI$ photometric and {\it I}-band time-series observations of the cluster NGC 2282 were carried out over 25 epochs during 15 Oct 2013 to 18 Feb 2016 using two different Indian telescopes. We performed photometric observations using  the $2048 \times 2048$  Andor CCD camera on the 1.3m Devasthal Fast Optical Telescope (DFOT) located at Nainital, India \citep[]{2011CSci..101.1020S}. The Andor CCD has a pixel size of 13.5 $\mu$m, which  provides a field-of-view (FOV) of about 18$\arcmin$ $\times$  18$\arcmin$ with a plate-scale of 0.535 arcsec pixel$^{-1}$ for F/4 optics of the DFOT. The Andor CCD gain is 6.5 e{$^-$}/Analog to Digital Unit(ADU), and  the readout noise is  2 e$^{-}$. The 2$\times$2 binning observing mode is used  to get relatively better signal to noise ratio (SNR). The typical FWHM of the stars were $\sim$ 2$\arcsec$ in our observing run. We also carried out photometric observations using Himalaya Faint Object Spectrograph Camera (HFOSC) on 2m Himalayan Chandra Telescope (HCT) located at Hanle, India \citep[]{2014PINSA..80..887P}.  The $2048 \times 2048$ HFOSC imaging CCD  provides a FOV of about 10$\arcmin$ $\times$  10$\arcmin$ with a plate scale of 0.296 arcsec pixel$^{-1}$ for F/9 optics of the HCT. The gain of the HFOSC CCD is 1.22 e{$^-$}/ADU, and  the readout noise is 4.8 e$^{-}$.  The imaging data is also taken with 2$\times$2 binning mode, and the typical FWHM of the stars were $\sim$ 1.5-3.2$\arcsec$. Long as well as short exposures are taken to get a good dynamical range.  The exposure time is decided in such a way that the brighter end is not saturated, and the fainter end has sufficient S/N (uncertainty <0.1). The log of our observations is shown in Table~\ref{tab:observation}.  Fig.~\ref{fig:optical_vri} shows the optical color composite image of NGC 2282, which was taken by the DFOT CCD camera having 18$\arcmin$ $\times$ 18$\arcmin$ FOV, and the HFOSC CCD FOV of 10$\arcmin$ $\times$ 10$\arcmin$ is marked for our overlapping monitoring program. 
     
The photometric reduction of the CCD frames were performed using IRAF\footnote{Image Reduction, and Analysis Facility (IRAF) is distributed by the National Optical Astronomy Observatories (NOAO), USA (http://iraf.noao.edu/)} software. Using various IRAF tasks, bias correction and flat correction were applied to remove the CCD characteristics and  cosmic rays from the raw images. IRAF DAOFIND task is used to detect the point sources in the science frames. Elimination of the extended sources like the background galaxies and brightness enhancements due to bad pixels from the point source catalog was done constraining the roundness limits of $-1$ to $+1$ and the sharpness limits of $0.2$ to $+1$ \citep[]{1987PASP...99..191S}. The ALLSTAR task of DAOPHOT package is used for PSF photometry  \citep[]{1992ASPC...25..297S}. The astrometry on the CCD frames was determined by the IRAF  $ccfind$, $ccmap$ and $ccsetwcs$ tasks using the coordinates of 20 isolated moderately bright stars from the 2MASS point source catalog (PSC) \citep[]{2003tmc..book.....C}. Positioning accuracy of better than 0.3$\arcsec$ was obtained.

The available optical to MIR archival  data are also used for characterizing the PMS stars of NGC~2282 such as  (i) optical $BVI$ data from paper I;  (ii) NIR $JHK$  data from the Galactic Plane Survey \citep[GPS;][data release 6]{2008MNRAS.391..136L}, which were obtained as a part of the UKIRT Infrared Deep Sky Survey \citep[UKIDSS,][]{2007MNRAS.379.1599L}; (iii) the  $JHK_s$ photometric data from the 2MASS PSC \citep[]{2003tmc..book.....C} for brighter end (Paper I); (iv) the {\it Spitzer}-IRAC observations at 3.6 and 4.5 $\mu$m bands (Program ID: 61071; PI: Whitney, Barbara A) from Paper I;  (v) MIR data at 3.4, 4.6, 12, and 22 $\mu$m from the Wide-field Infrared Survey Explorer (WISE) All-sky Survey Data Release \citep[]{2010AJ....140.1868W}; (vi) the optical data  in Sloan broad $r$, $i$  and narrow-band $H\alpha$ filters were obtained from the INT Photometric $H\alpha$ Survey of the Northern Galactic Plane (IPHAS), which was conducted using the Wide Field Camera (WFC) on the 2.5m Isaac Newton Telescope (INT) \citep[]{2005MNRAS.362..753D,2008MNRAS.388...89G,2014MNRAS.444.3230B}.

\begin{figure*}
 \includegraphics[width=8 cm,height=6.0cm]{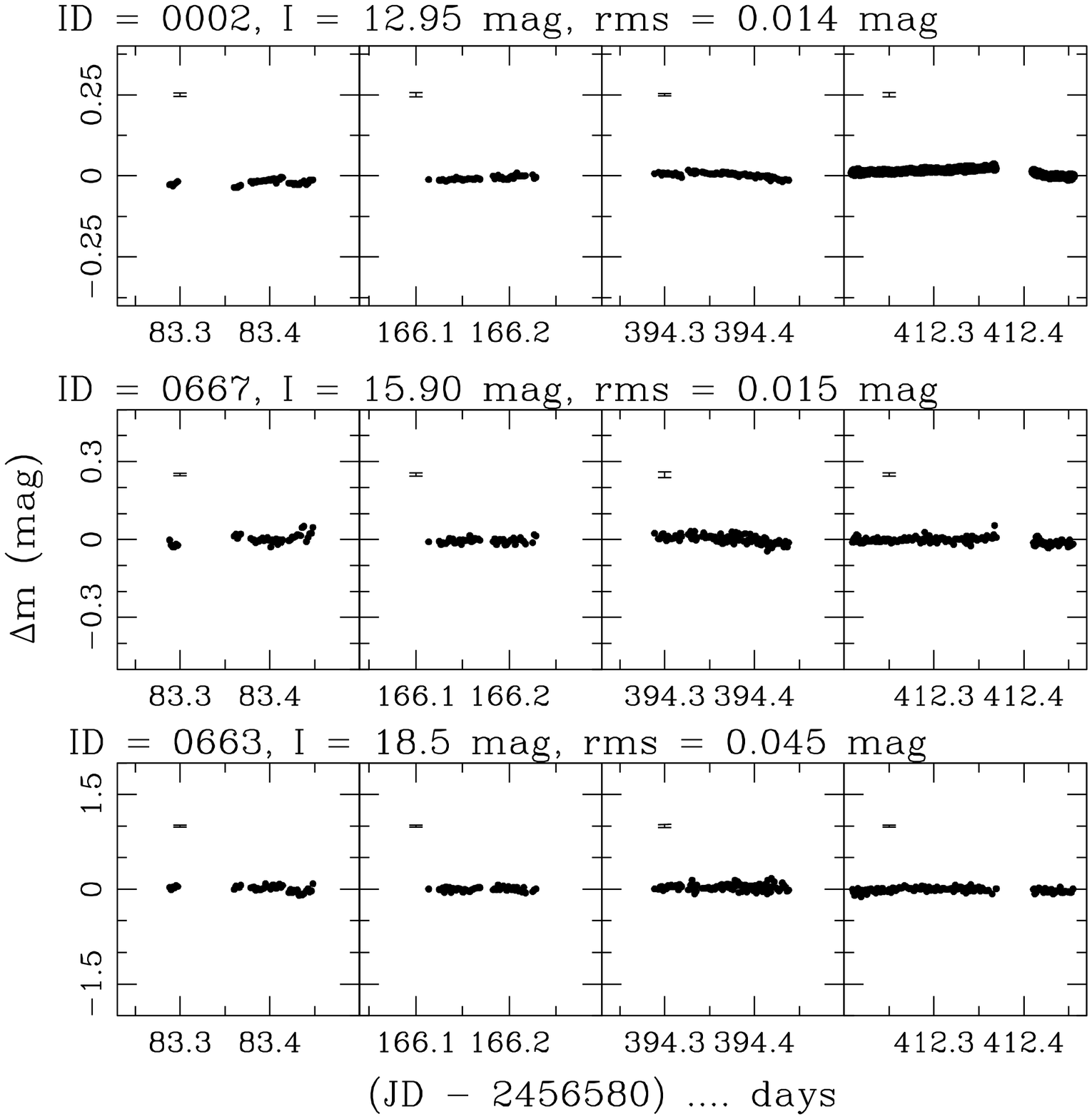}
 \includegraphics[width=8 cm,height=6.0cm]{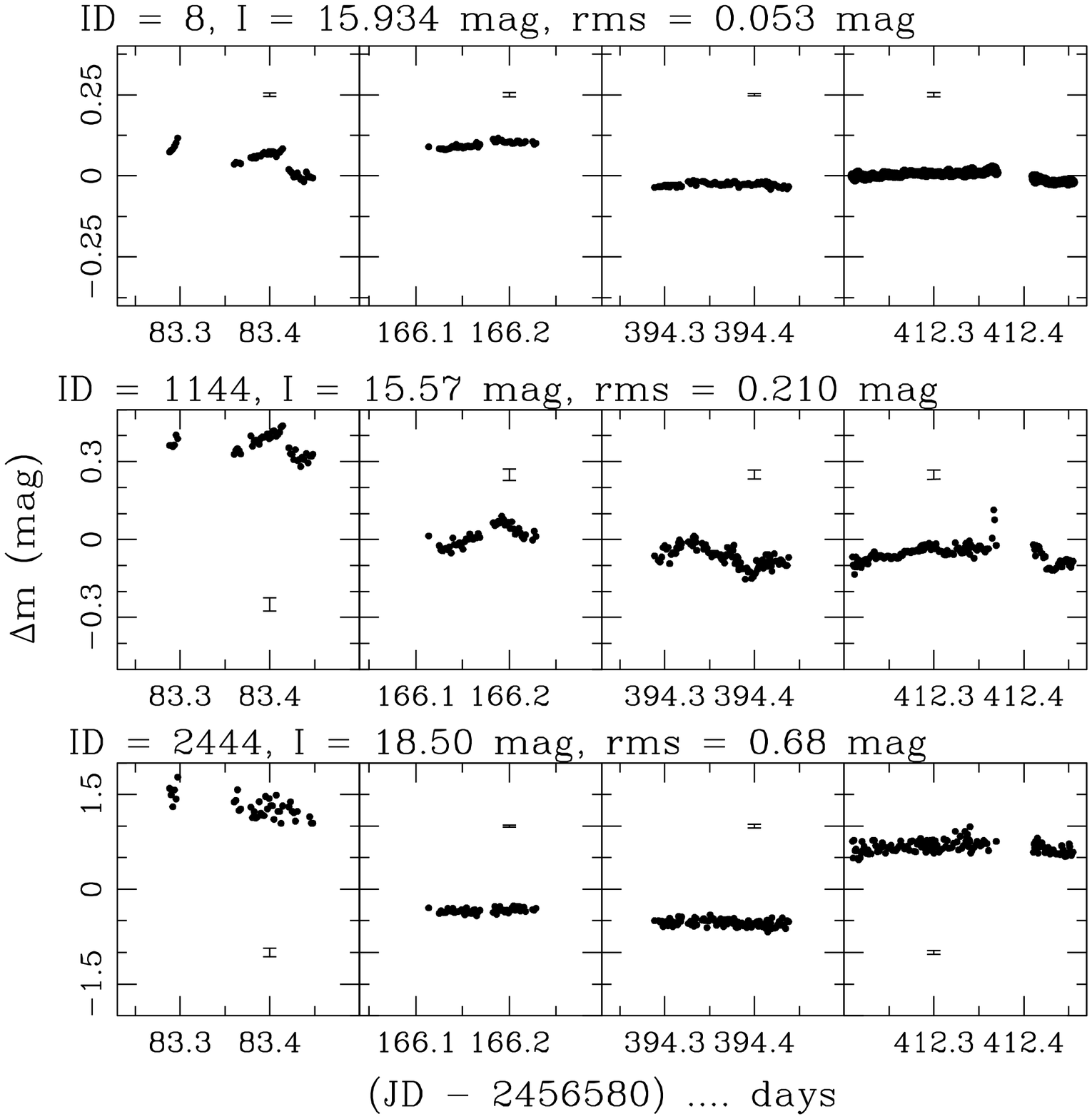}
  \caption{Example light curves for a few apparently non variable stars of different magnitude range (left three panels) and a few candidate variable stars (right three panels). The $\Delta$m represents the relative magnitude with respect to  reference star. Star identifications, $I$ magnitude and RMS value are mentioned. Average error corresponds to each day data is plotted (see text for details).}
  \label{fig:diff}
\end{figure*}

\section{Results and discussion}

\subsection{Identification of variable stars}
\subsubsection{Differential Light Curves}

Differential photometry was performed to clean the light curves from sky variability, instrument signatures, airmass, etc.  We considered a few reference stars in the same CCD frame of the targets, which showed a stable behavior from root-mean-square (RMS) deviation as discussed later and visual inspection of the observed light curve. To obtain relative photometry, we subtract each target star magnitude from those of the reference stars in each frame. The technique of differential photometry provides excellent photometric precision, even in the crowded nebulous region of the cluster. The light curves of a few non-variable and variable stars are displayed in Fig.\ref{fig:diff}. The catalog of detected variable stars from our analysis are listed in Table~\ref{tab:cat_var}.

\begin{figure}
 \includegraphics[width=8 cm,height=6.0cm]{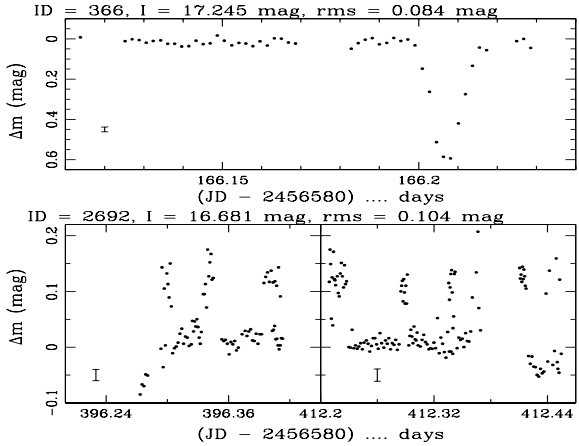}
  \caption{Example light curves of a eclipsing binary (upper panel) and probably fast stochastic variable  (lower panel). All the marks are as Fig.~\ref{fig:diff}  (see text for details).}
  \label{fig:delta_w}
\end{figure}

Two interesting example light curves are shown in Fig.~\ref{fig:delta_w}. From the light curves analysis, we identified ID 366 as an eclipsing binary star. Another star ID 2692 shows fast stochastic variability, could be a classical T Tauri candidate as a typical manifestation of the underlying variable accretion processes. 

\subsubsection{RMS analysis}

Considering a large number of sample, one quantitative estimator of photometric variability, i.e., the RMS deviation in the differential photometric light curves is utilized. The mean value of the instrumental magnitudes and the RMS deviation in the light curves were estimated for each star over the whole span of observations for both the telescopes to select the variable stars. Following \citet[]{2001AJ....121.3160C},  the observed RMS  ($\sigma_{mag}$) for each star was estimated using the following relation as:

\begin{equation}
\sigma^2_{mag} = \frac{n \sum_{k=1}^{n} w_k (I_k - \bar{I})^{2}}{(n - 1) \sum_{k=1}^{n} w_k},
\end{equation}
where $I_k$'s are the individual magnitudes associated with photometric uncertainty $\sigma_k$'s and weightage $w_k = 1.0/\sigma_k^2$ for each observation.

\begin{figure}
 \includegraphics[width=8 cm,height=6.0cm]{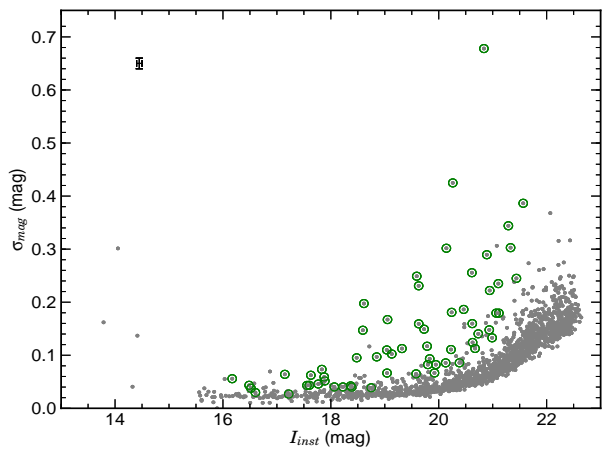}
  \caption{RMS ($\sigma_{mag}$) of the $I_{inst}$-band light curves of 1627 stars towards NGC 2282. The green circles are the candidate variables. The average error bars for magnitudes and rms value are also shown on the top left corner (see text for details).}
  \label{fig:rms}
\end{figure}

Fig.\ref{fig:rms} displays the RMS of each star as a function of mean instrumental magnitudes. The RMS of majority stars increases with fainter magnitudes as the signal-to-noise ratio (SNR) decreases towards the fainter limit. The observed RMS values of normal stars follow an expected nearly exponential trend; the RMS values range from $\sim$ 0.01 mag for a brighter end to $\sim$ 0.2 mag towards completeness limit. However, a few stars are scattered from the normal trend. Such type of deviation could be due to the photometric noise or intrinsic variability.  From outliers objects in  Fig.\ref{fig:rms}, we picked up variable candidates, which are showing RMS equal or more than three times the mean RMS at each half-magnitude bin. We selected 69 variable stars from such criterion. Visual inspection of the light curves indicated that a few of them show large RMS values because of only a few data points, which may be the results of bad pixels, cosmic rays and falling on the edge of the CCD. We rejected those from variable list. Finally, we considered 62 variable candidates for further analysis.

The star-forming history of the young clusters could be probed from the spatial distribution of variable PMS stars. Fig.~\ref{fig:optical_vri} displays the spatial distribution of the candidate variables towards NGC 2282 (white circles). Majority of the variables are concentrated in the core region of the cluster in the  Fig.~\ref{fig:optical_vri}, while a significant scattered population is seen towards the north-east part. Such distribution suggests that the cluster might be extended towards that direction as predicted by \citet[]{1997AJ....113.1788H}.

\subsubsection{Period Analysis}

\begin{figure}
 \includegraphics[width=8 cm,height=6.0cm]{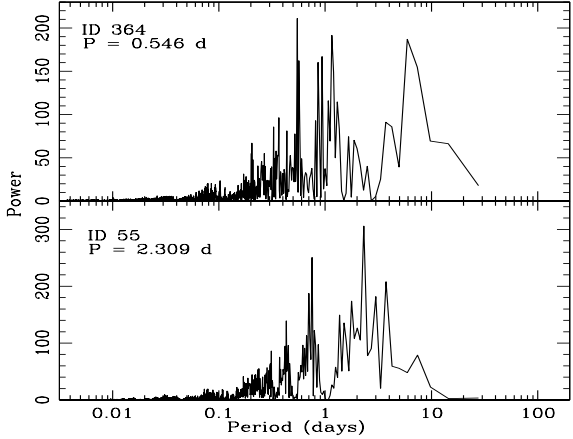}
  \caption{Example LS power spectrum of selected stars. The highest peaks are considered for period estimation (See text for details).}
  \label{fig:freq}
\end{figure}

\begin{figure}
 \includegraphics[width=8 cm,height=5.0cm]{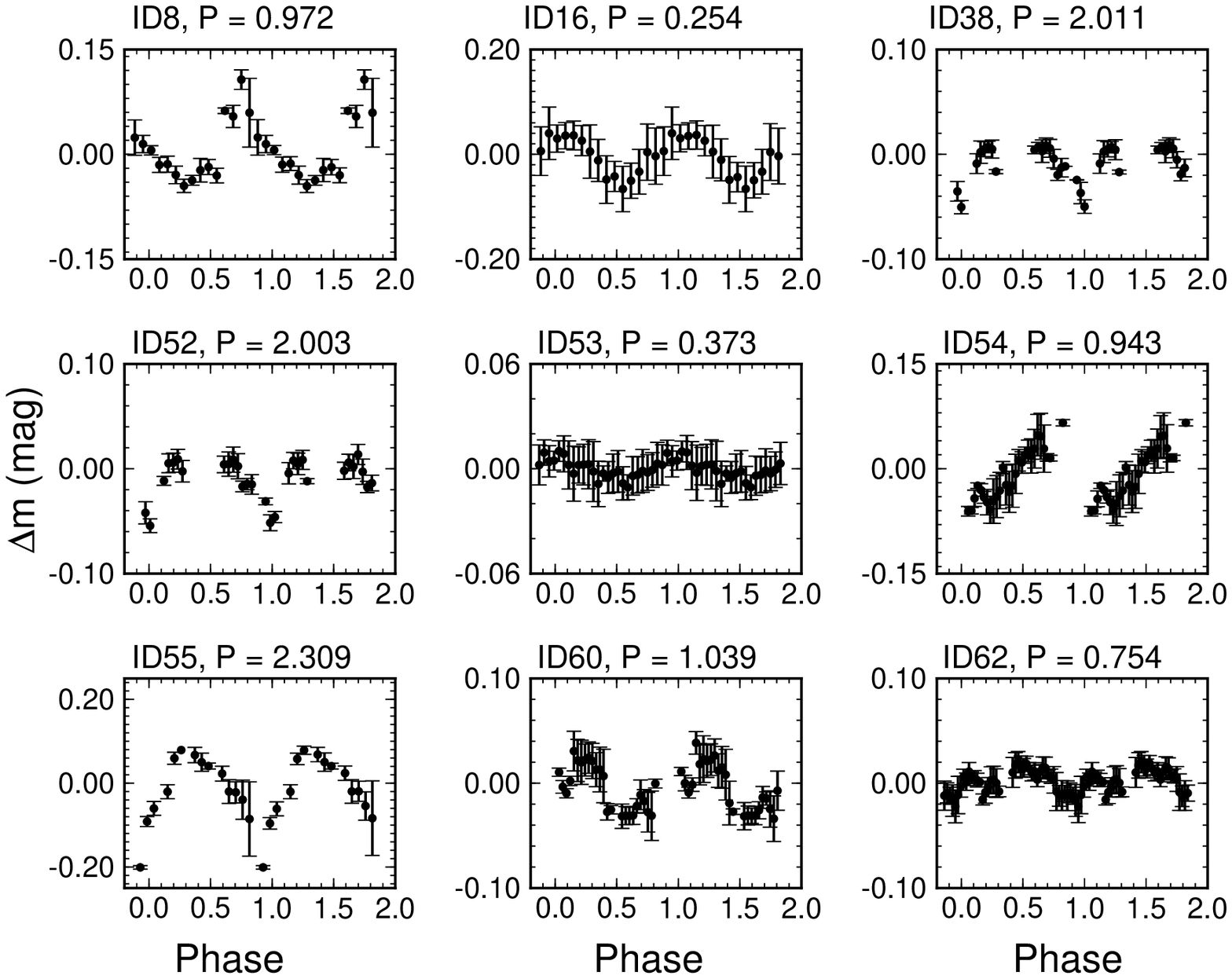}
 \includegraphics[width=8 cm,height=5.0cm]{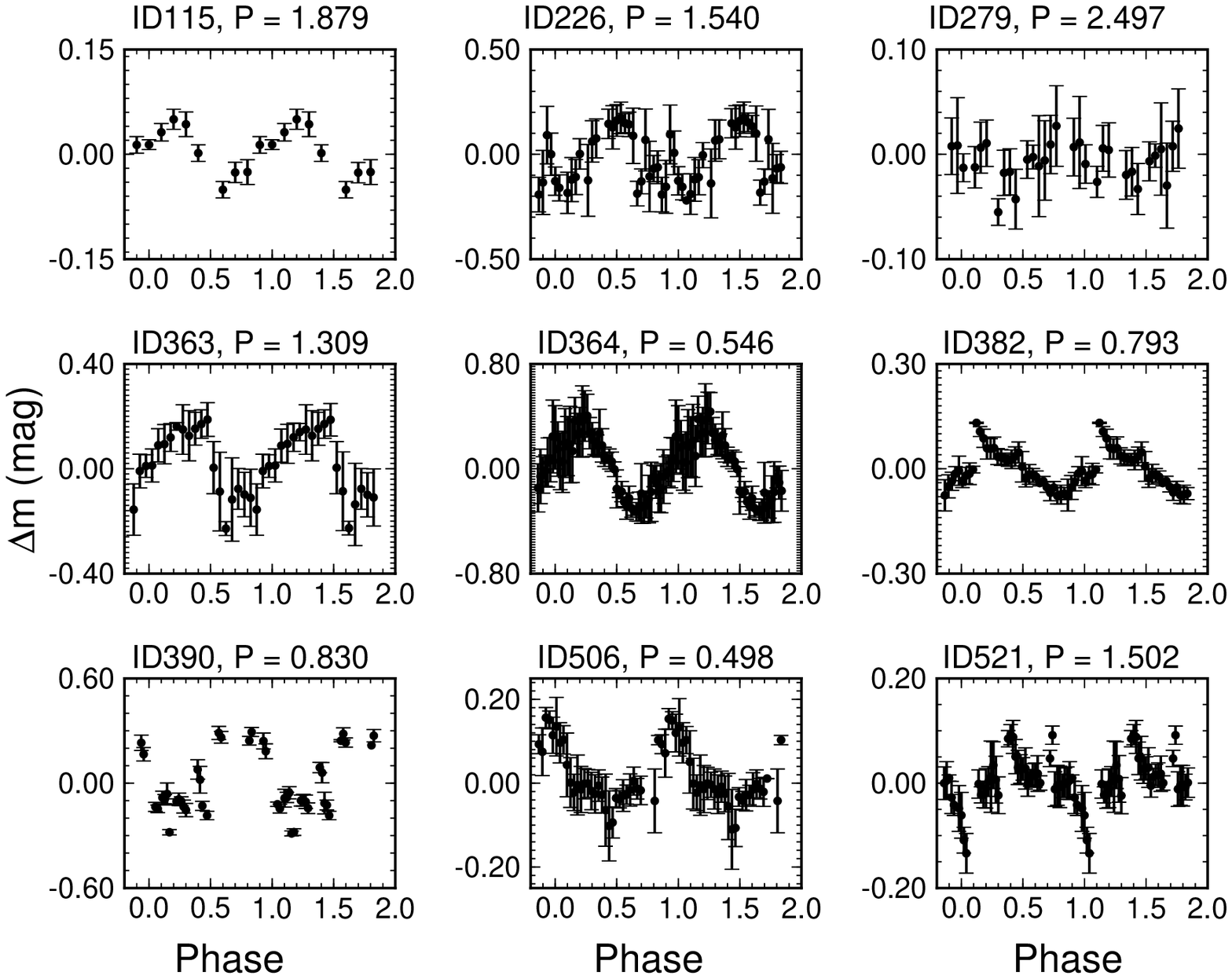}
 \includegraphics[width=8 cm,height=5.0cm]{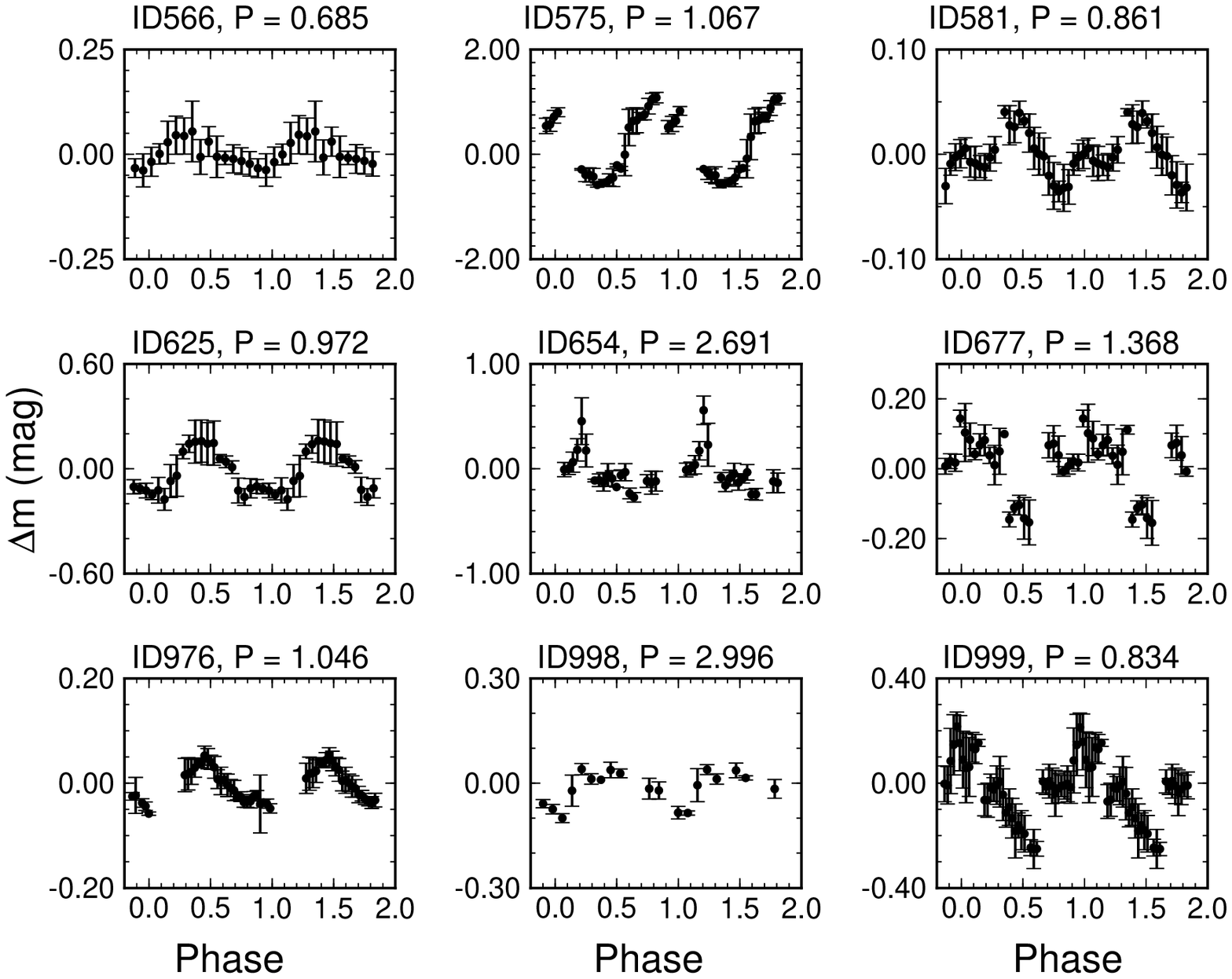}

  \caption{The phase light curve of variables are presented here. Period of variables are taken from Table~\ref{tab:var_phy}, and the details of the variable IDs are listed in Table~\ref{tab:cat_var} (see text for details).}
  \label{fig:phase}
\end{figure}

The Lomb-Scargle (LS) periodogram analysis \citep[]{1976Ap&SS..39..447L,1982ApJ...263..835S} is used to estimate periods of those variable candidates. The LS algorithm is publicly available with the Starlink\footnote{http://www.starlink.uk} software package. The LS method is used to find out significant periodicity even with unevenly spaced observed data, and verified successfully in several cases to determine periods from sparse data sets \citep[]{2004A&A...417..557L,2010AJ....139.2026M,2012MNRAS.427.1449L}. The periods were also verified from other software like PERIOD04\footnote{http://www.univie.ac.at/tops/Period04}, which provides the frequency as well as semi-amplitude of the periodic light curve \citep[]{2005CoAst.146...53L}. The periods were further verified by online LS periodogram analysis available on NASA Exoplanet Archive Periodogram Service \footnote{http://exoplanetarchive.ipac.caltech.edu/cgi-bin/Periodogram/nph-simpleupload}. Out of 62 variable candidates, 41 stars are identified as periodic variables. The LS power spectrum of two variable candidates with identification numbers ID 364 and ID 55 are shown as an example in Figure~\ref{fig:freq}, which show significant periods of 0.546 and 2.309 days, respectively. The significant periods were always verified with visual inspection of the phase light curves and accepted as periodic variables. The periods of variable candidates exhibit in the range 0.233 to 7.143 days and are listed in Table~\ref{tab:var_phy}. Considering those estimated periods, we generated the phased light curves of variable candidates, and they are presented in Fig.~\ref{fig:phase}.

\subsection{Characterization of variable stars}

\subsubsection{H$\alpha$ Emission property of Variable candidates}

\label{sec:h_alpha}

The H$\alpha$ emission indicates the young stellar accretion activity in T Tauri phase. Hence this phenomenon could be used to identify the candidate members of the cluster NGC 2282. We made use of the IPHAS archive data as described in section 2 to identify the $H\alpha$ emission variable stars among the variable candidates (see also Paper I). Fig.~\ref{fig:barentsen} presents the IPHAS ($r-i/r-H\alpha$) CC diagram of all the candidate variable stars.  The nearly vertical green and black lines are the trend of an unreddened optically thick disk continuum and an unreddened Rayleigh-Jeans continuum, respectively \citep[]{2014MNRAS.444.3230B}. The predicted emission lines with constant EW are shown as the black broken lines.  The locus of unreddened main-sequence is represented by the blue lines (solid) and the main-sequence with $-$10 \AA $~$ H$\alpha$ emission line (broken), respectively.   The stars located above 3$\sigma$ of the main-sequence line having EW $-$10 \AA $~$ are considered as Classical T Tauri Stars (CTTSs).  However, the identification of CTTSs solely from IPHAS photometry poses difficulty due to the presence of other possible H$\alpha$ emission objects like Be stars, luminous blue variables, interacting binaries, Mira variables, unresolved planetary nebulae, etc.

We identified 21 variable sources as the $H\alpha$ emitting stars from the IPHAS photometry following method as mentioned in the above. We also compared $H\alpha$ emission properties from the IPHAS photometry to that of slitless spectroscopy (Paper I). Rest of the variable stars could be WTTSs with small $H\alpha$ or no $H\alpha$ emission.  However, variable $H\alpha$ emission might be a possible reason for such non-detection, even though they emit strong $H\alpha$. The details of the variable candidates having $H\alpha$ emission in NGC 2282 region are listed in Table~\ref{tab:var_phy}.

\subsubsection{NIR Color-Color Diagram}

Like H$\alpha$ emission, the NIR excess is a useful diagnostic technique to identify the PMS stars, and such technique is used here to establish the variable candidates as probable PMS objects.  We plotted NIR $(J - H)/(H - K)$ CC diagram of the  variable candidates in Fig.~\ref{fig:jhk_var} using $JHK$ data sets as described in section 2. The black solid and long dashed red lines represent the locus of the intrinsic colors of dwarf and giant stars \citep[]{1988PASP..100.1134B}. The dashed black line is the locus of the CTTSs \citep[]{1997AJ....114..288M}.  The intrinsic locus and observed photometric data were transformed into the CIT system following the relations are given by \citet[]{2001AJ....121.2851C}. The different interstellar reddening vectors (i.e., $A_J$/$A_V$ = 0.265, $A_H$/$A_V$ = 0.155 and  $A_K$/$A_V$ = 0.090) are shown as parallel dashed blue lines, which were adopted from \citet[]{1981ApJ...249..481C}. 
The $JHK$ NIR space is divided into F, T, and P sub-regions. The NIR emission of stars in  `F' region is believed to originate mainly from their bare photosphere; they might be either the field stars or the WTTSs/Class III sources with small or no NIR excess. The NIR CC diagrams are not sufficient to distinguish between the WTTSs with small NIR excess and the field stars \citep[]{2004ApJ...608..797O}. The stars located in the  `T' region could be considered as the CTTSs, and they possess optically thick accretion disk \citep[]{1997AJ....114..288M}. This excess emission arises from both photosphere and circumstellar disk \citep[]{1992ApJ...393..278L}. The `P' region stars are considered as protostars, and more NIR color excess at K-band might be originating from the thick envelope or very thick accreting disk around them \citep[]{2012ApJ...755...65R}. 

From Fig.\ref{fig:jhk_var}, we find that majority of the variable stars follow the CTTSs locus.  Out of 62 stars, 30 stars (47\%) were previously identified as the young stars from either infrared excess, H$\alpha$ emission or both activities. The remaining 32 of 62 variables might be the WTTSs with less IR excess or field stars, and they are located at `F' region in Fig.~\ref{fig:jhk_var}. But spectroscopy follow up is needed to distinguish them from the field stars.  However, the YSOs selection in Paper I is still incomplete because of non-availability of longer wavelength data in highly obscured nebulous region.

\addtocounter{figure}{-1}

\begin{figure}

  \includegraphics[width=8 cm,height=5.0cm]{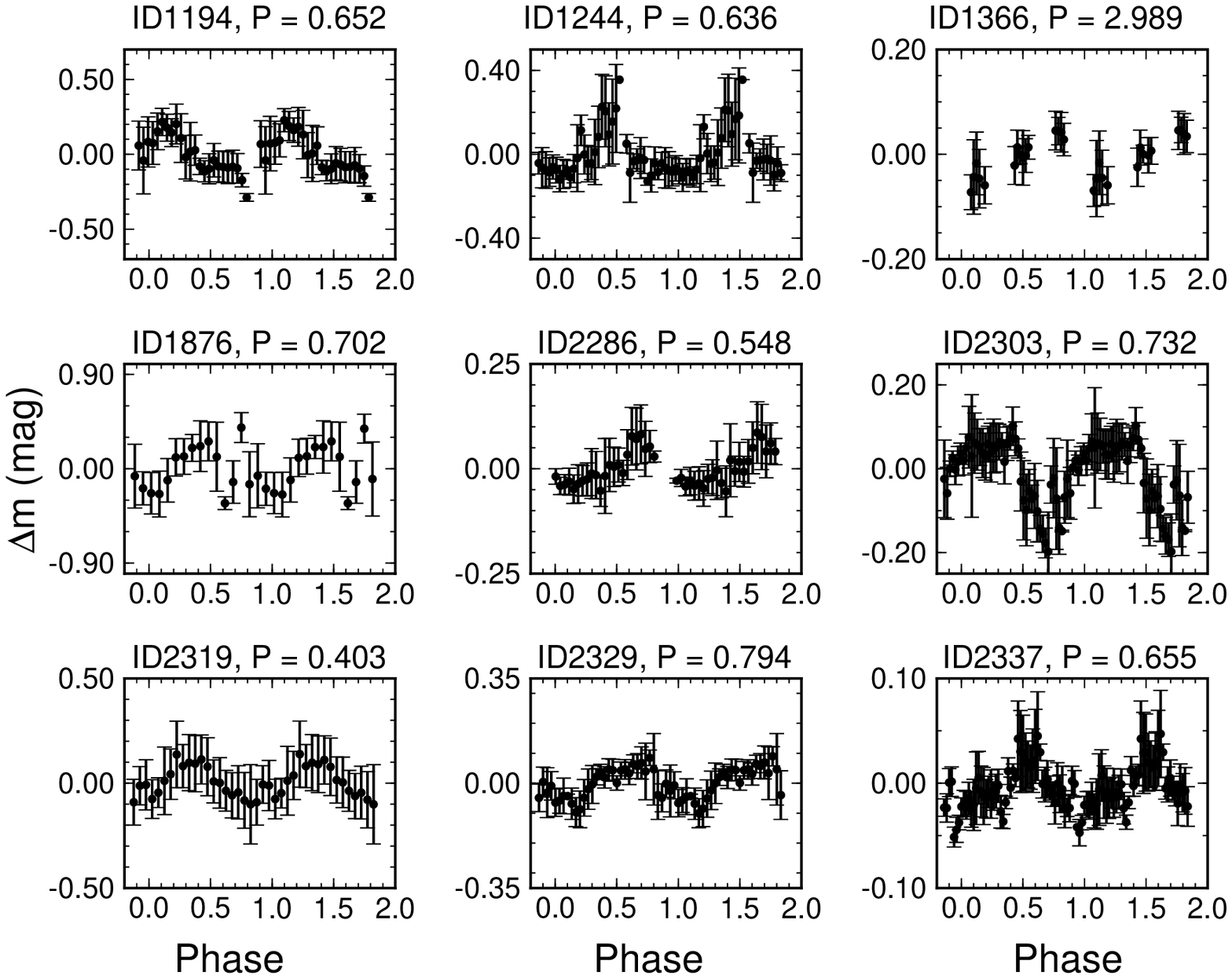}
  \includegraphics[width=8 cm,height=3.7cm]{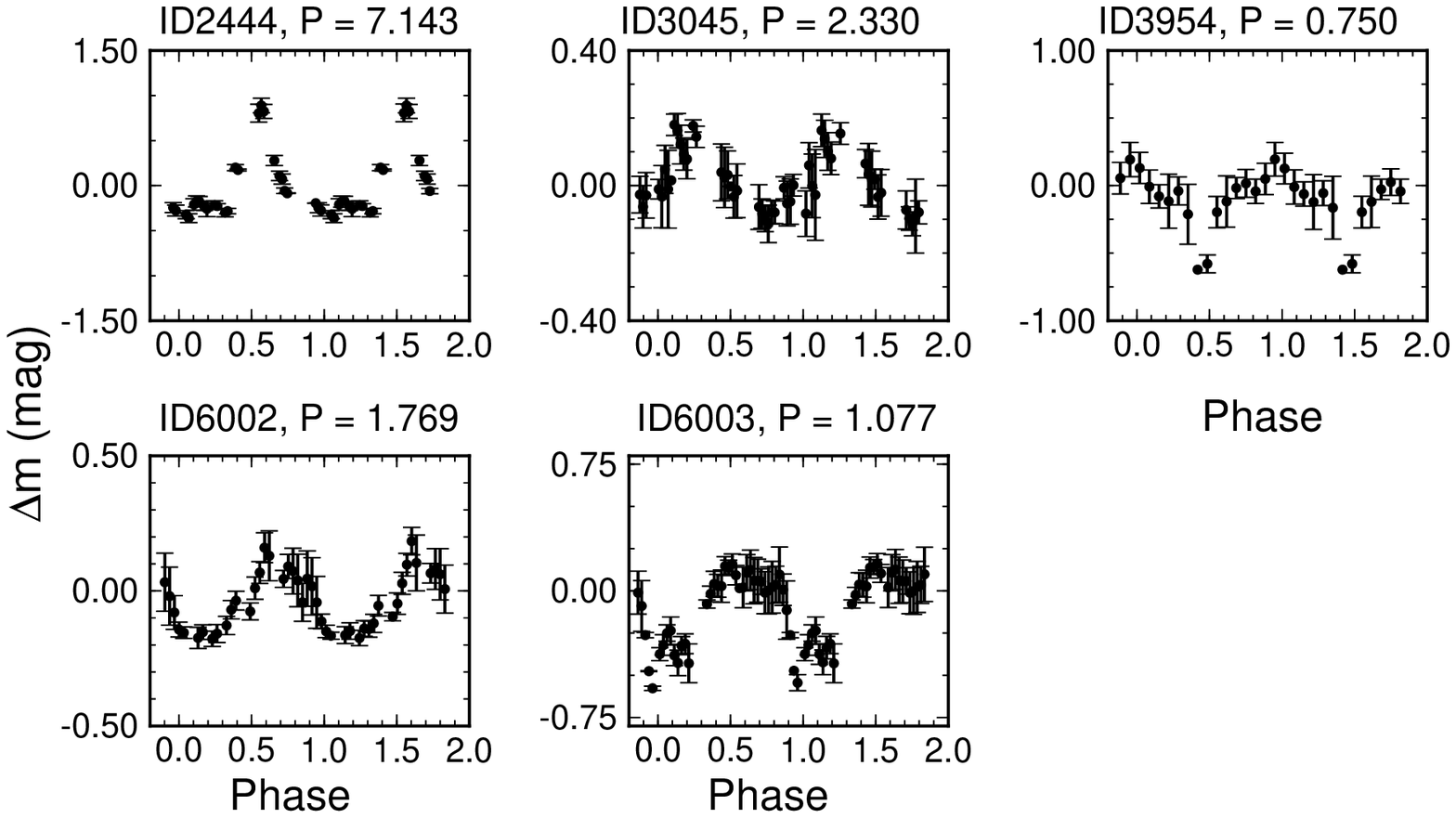}
  \caption{Same as previous figure.}
  \label{fig:phase1}
\end{figure}

\begin{figure}
\includegraphics[width=8.0 cm,height=8.0 cm]{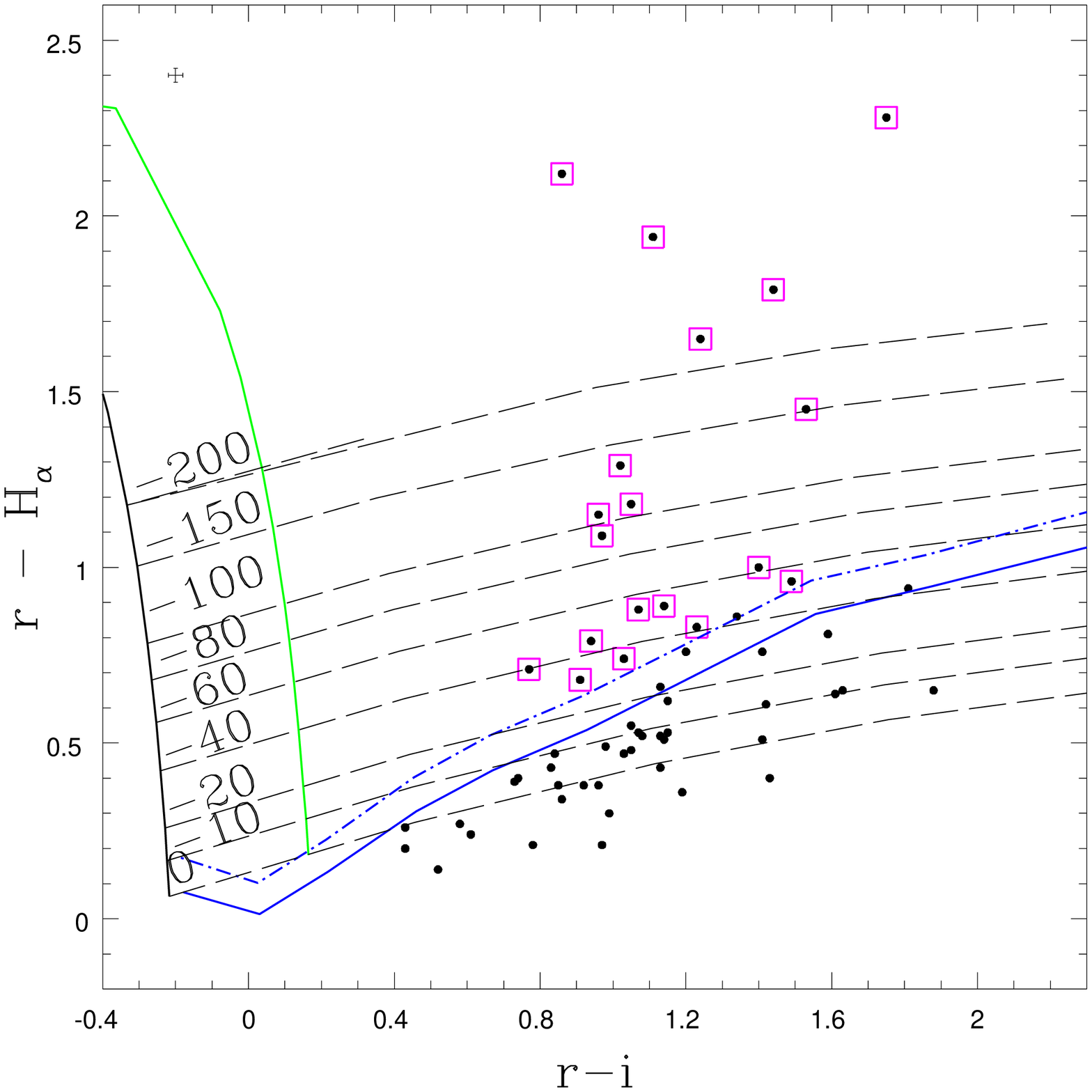}
 \caption{($r-i$) versus ($r-H\alpha$) CC diagram for all the variable sources detected in the IPHAS catalog towards NGC 2282. The average error bars are also shown on the top left corner.  The magenta boxes are the  identified H$\alpha$ emission sources from the IPHAS photometry. See text for unreddened continuum and locus of main-sequence and increasing levels of $H\alpha$ emission tracks.} 
 
  \label{fig:barentsen}
\end{figure}

\begin{figure}
\includegraphics[width=8.0 cm,height=8.0 cm]{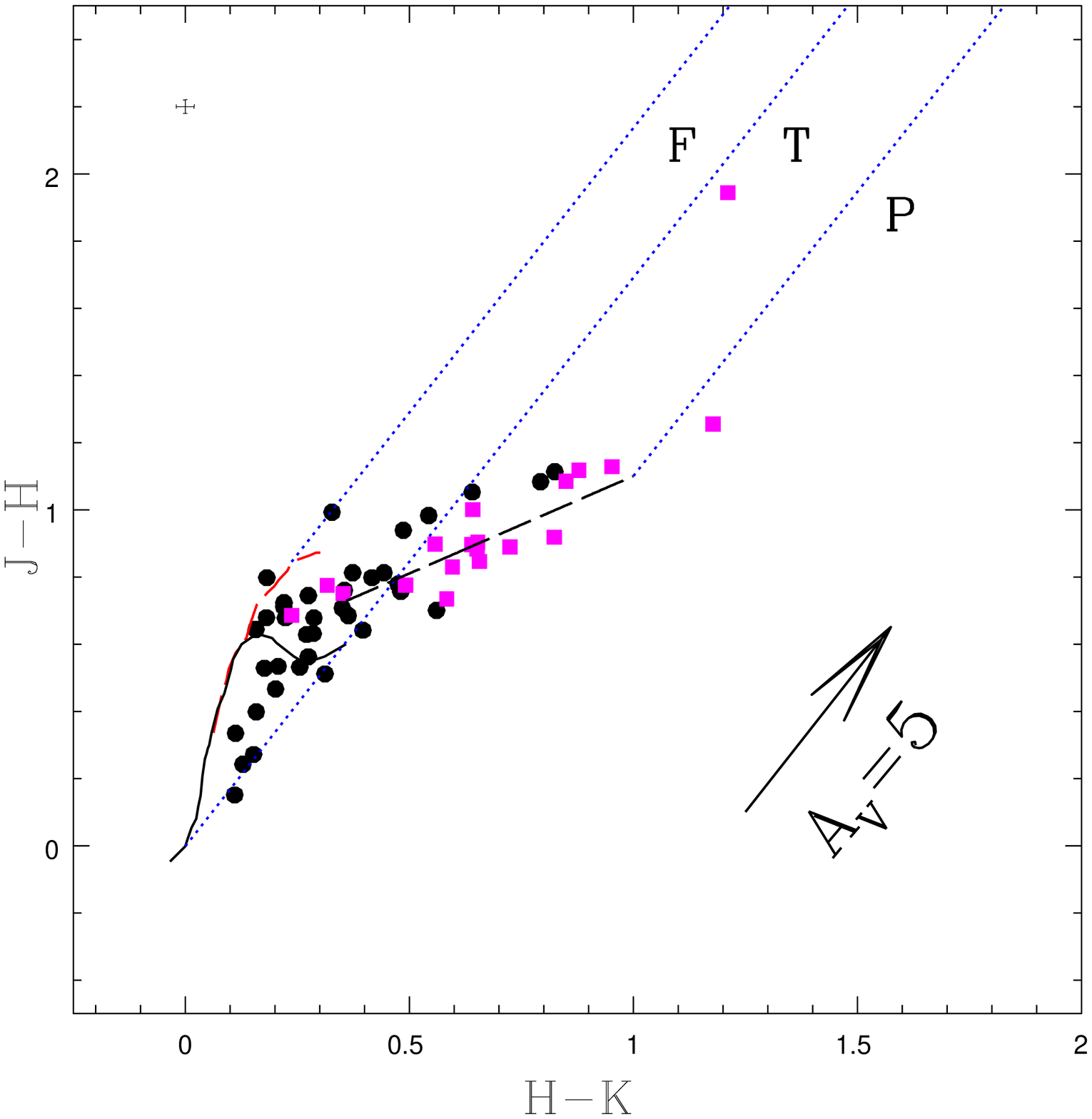}

\caption{The ($J-H$) versus ($H-K$) CC diagram of variable stars within 3.15$\arcmin$ radius.  The magenta filled squares are H$\alpha$ emission objects detected in Fig.~\ref{fig:barentsen} and the black filled circles are non-emitting objects.   The locus of dwarf (black solid), giants (red long-dashed), CTTSs (black long dashed) and the reddening vectors (blue small dashed) are shown here. The locus for dwarfs and giants  are taken from \citet[]{1988PASP..100.1134B}, while the locus of CTTSs are adopted from \citet[]{1997AJ....114..288M}. The reddening vector of visual extinction $A_V$ = 5 mag is also marked. The typical error bars are marked on the top left corner.}
  \label{fig:jhk_var}
\end{figure}

\subsubsection{Optical Color-Magnitude Diagram of YSOs}

\begin{figure}
\includegraphics[width=8.0 cm,height=8.0 cm]{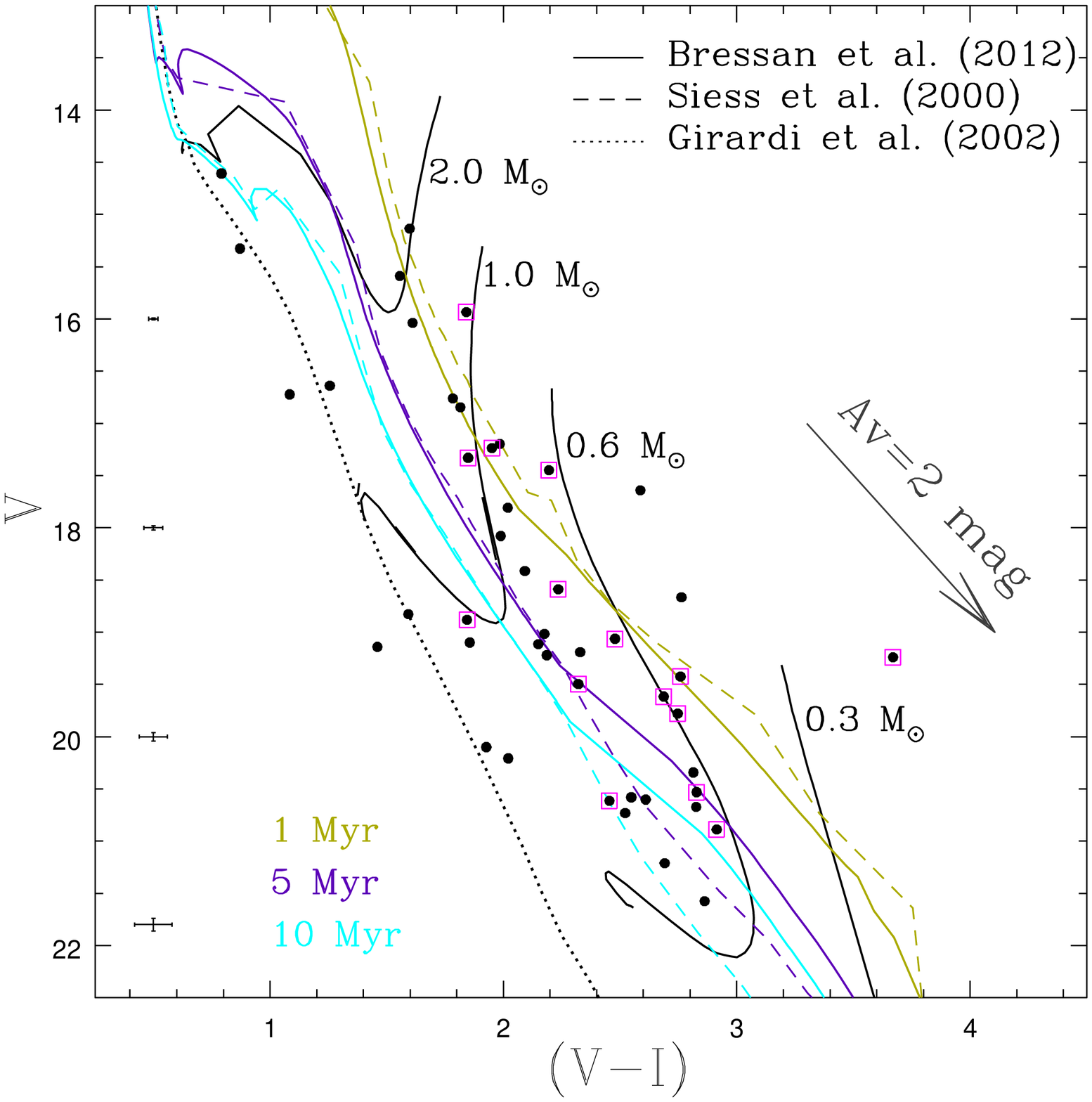}
  \caption{V/(V$-$I) CMD for variable stars towards NGC 2282. All symbols are same as in Fig.~\ref{fig:barentsen}.  The locus of ZAMS adopted from \citet[]{2002A&A...391..195G} is shown as the dotted curve, while the solid lines are the PMS isochrones of age 1.0, 5.0 and 10.0 Myr, respectively. The evolutionary tracks for various mass bins adopted from \citet[]{2012MNRAS.427..127B} are shown as the thin black solid lines. The average errors in $V$ and ($V-I$) colors are displayed on the left side of the figure.}
  \label{fig:viv_yso}
\end{figure}

The ages and masses of the cluster members could be approximately estimated using the optical CMDs. In Fig.~\ref{fig:viv_yso}, we plotted $V$ vs $(V-I)$ of the PMS variables using the $VI$ data sets from Paper I (see section 2).   In Fig.~\ref{fig:viv_yso}, the locus of  ZAMS adopted from \citet[]{2002A&A...391..195G} is shown by the dotted curve. The ages and masses of the PMS variables are estimated from the PMS isochrones and evolutionary tracks taken from \citet[]{2012MNRAS.427..127B}, which is plotted in Fig.~\ref{fig:viv_yso}. The PMS isochrones from \citet[]{2000A&A...358..593S} are also shown for comparison. The isochrones and evolutionary tracks were corrected considering the cluster distance 1.65~kpc and reddening of $E(V-I)$ = 0.65 mag ($E(B-V)$ = 0.52 mag) (see Sect. 3.3 \& 3.4 of Paper I). The reddening vector in CMD space is nearly parallel to the theoretical isochrones. Therefore, a small change in extinction would not show a significant change in the age estimation of the PMSs.

We estimated the masses and ages of the PMS stars using interpolation methods of isochrones and tracks in the CMD compiling $VI$ photometry. Our estimation is limited due to the absence of $V$ measurements of all variable stars. The position of  PMS variables in the CMD  is spread over 1$-$10 Myr. However, different models at the low-mass end differ significantly as seen in Fig.~\ref{fig:viv_yso}. The average mass of the YSOs seems to be $\sim$ 0.3$-$2.5 M$_\odot$.

\subsection{Period distribution}

\begin{figure}
\includegraphics[width=8.5 cm,height=8.0 cm]{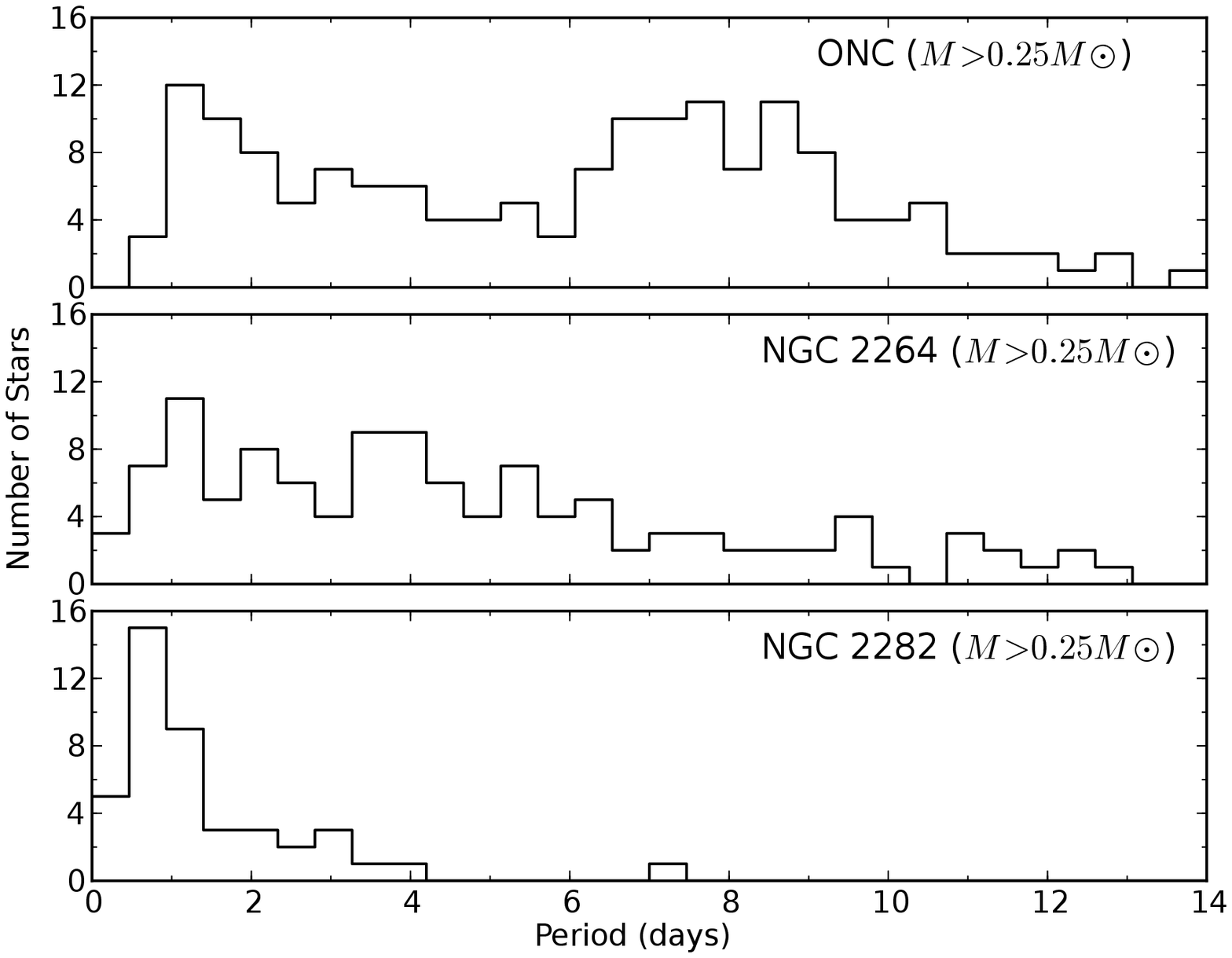}

\caption{The distribution of periods in NGC 2282 is compared with that of  the ONC \citep[]{2002A&A...396..513H} and NGC 2264 \citep[]{2004AJ....127.2228M}. The bimodal period distribution is seen in ONC and NGC 2264, while the distribution for NGC~2282 is uni-modal.}
  \label{fig:comp}
\end{figure}

All variable stars are searched for periodicity; however, we were managed to estimate periods of $\sim$ 66\% stars. Figure~\ref{fig:comp} displays frequency distribution of rotation periods in NGC 2282. It is to be noted that we have a large number of short-period  variables with about 50\% and 75\% objects having periods  less than 1  and 2 days, respectively. We have less number of relatively long (more than 2 days)  periodic variables. There are no particular biases for slow or fast rotation periods in our analysis. But, uneven sampling data with long gaps in successive observations without any long single stretch of observations might be a reason to detect a low number of relatively long periodic variables. Again, periodic variability mainly due to spots, which evolve with time and a long gap in data might have masked out such long periods. The period distribution of NGC~2282 is compared with Orion nebula cluster (ONC) stars and NGC 2264 stars with masses greater than 0.25 M$_\odot$.  \citet[]{2000AJ....119..261H,2002A&A...396..513H} found a bimodal distribution of the periods for the  stars $>$ 0.25 M$_\odot$ in ONC. But no such conclusive evidence was found for NGC 2264 cluster \citep[]{2004AJ....127.2228M}.  However, the faint low-mass objects are not covered in their study as NGC 2264 is located at more far distance than ONC. For NGC 2282 stars, the period distribution is unimodal, where fast rotators are peaking up at $\sim$ 0.5-1 days. 
\citet[]{2002ApJ...566L..29H} suggests that the disk locking in highly accreting protostars with smaller magnetosphere have faster rotation periods, and they spin down gradually at the T Tauri phase over the disk braking timescale, which implies that the disk-locking may not be efficient at very early stages in protostars. \citet[]{2002A&A...396..513H}, performed a detailed study on ONC, argued that the stars are released from their disk locking after $\sim$ 1 Myr of age. However, \citet[]{2005MNRAS.356..167M} argued  that the time scale of disk-locking should be $<$ 4 $\times 10^4$ yr. After a comparative study on ONC, Orion Flanking Fields and IC 348, \citet[]{2005A&A...437..637L} concluded that disk locking takes effect at $\sim$ 1 Myr, and at the age of $\sim$ 3 Myr (IC 348) disk locking is still effective. However, the disk-locking theory could not explain the notable difference of period distribution for NGC 2264 and IC 348. The period distribution of NGC 2282 is strongly biased towards fast rotation side and the lack of slow rotators is quite significant. Our sample mainly consists of stars with periods below four days. This contradiction can be explained by the fact that the size and luminosity of a young star increases dramatically during their rapid accretion phase \citep[e.g.][]{1996ARA&A..34..207H,1999ApJ...518..833K}. Hence, the star would appear much younger in the CMD than it's non-accreting counterpart of the same radius and initial mass \citep[]{2011MNRAS.413L..56L}. However, some of the results of such period distribution may be effected from cluster environments like variability, extinction, and binarity, etc. \citep[]{2005A&A...437..637L,2011MNRAS.413L..56L}.

 \subsection{Correlation between IR excess and rotation periods and variability amplitude}
\label{correlation_IR_rot_amp}

From the time-series photometric observational data on the young (1-5 Myrs) star-forming regions, several authors have investigated any possible correlations between 
the rotational periods and the presence of circumstellar disks in PMS stars using IR excess and H$\alpha$ emissions. The results reveal  significant differences in the rotation periods between stars with circumstellar disks and those lacking disks, and others studies do not find any conclusive evidence \citep[e.g.][]{2001AJ....121.1676R,2002A&A...396..513H,2002AJ....124.1001C,2004AJ....127.2228M,2005A&A...430.1005L,2006ApJ...646..297R,2007ApJ...671..605C,2009A&A...508.1301B,2009ApJ...695.1648N,2010A&A...515A..13R,2010ApJS..191..389C,2015AJ....150..132R}.

\subsubsection{Disk identification}

To draw any possible correlation of the NIR excess with rotation rates of our periodic/aperiodic variables, we invoked an index ($I-K$) as defined by \citet[][]{1998AJ....116.1816H} and  subsequently used by several authors \citep[e.g.][]{2002A&A...396..513H,2004AJ....127.2228M,2010A&A...515A..13R}. The $I$-band fluxes are believed to be dominated by pure photospheric emission, whereas the $K$-band fluxes have additive circumstellar disk emission on the photospeheric emission \citep[][]{1998AJ....116.1816H,2010A&A...515A..13R}. The ($I-K$) index being longer wavelength base provides more pronounce excess compared to other NIR indices (e.g. ($J-H$) or ($H-K$)). The NIR excess from ($I-K$) color defined by \citet[][]{1998AJ....116.1816H} and the same used by others \citep[e.g.][]{2002A&A...396..513H,2004AJ....127.2228M,2010A&A...515A..13R}, can be written as,

\begin{equation}
\label{eq:ik}
\Delta(I-K) = (I-K)_{obs} -(A_I - A_K) - (I-K)_o
\end{equation}

Where is ($I-K$)$_{obs}$ is the observed magnitudes; A$_I$ and A$_K$ are the interstellar extinctions in $I$ and $K$-bands, respectively. ($I-K$)$_o$ is the intrinsic color of the star. Without knowledge of spectral types, the intrinsic color is difficult to measure, and this is the case for our variables. However, it could be estimated from photometric measurements, extinction, and theoretical isochrones. Following \citet[]{2010A&A...515A..13R}, first, we measured the intrinsic colors ($I-J$)$_o$ as it is expected to be originating from photospheric emission from observed ($I-J$) colors corrected by the average extinction of A$_v$ = 1.65 mag (see Paper I). Using the \citet[]{1998A&A...337..403B} color information of NGC~2282 and the mean age of 3 Myrs (Paper I), we estimated the intrinsic color ($I-K$)$_o$ from ($I-J$)$_o$.  The NIR excess $\Delta(I-K)$ are calculated using Eqn.~\ref{eq:ik} for individual variables. 
It is to be noted that intrinsic color estimation of the young stars by comparing the observational data with the theoretical isochrones of mean age corrected for average extinction is an unrealistic approach, which includes several errors from nonuniform extinction, excess disk emission, variability, and binarity. However, it provides indicative results for further observations.

To search possible correlation between the rotation rates and the NIR excess we have plotted period versus  $\Delta(I-K)$. Following \citet[]{2002A&A...396..513H,2004AJ....127.2228M,2010A&A...515A..13R}, the value of $\Delta(I-K)$ = 0.3 mag is adopted for separating the disk and disk-less stars. Out of 41 periodic variables, we don't have K-band measurements for two objects. We plotted 39 periodic variables in the upper panel of Fig.\ref{fig:per_IRexcess} (upper panel), where 16 objects show NIR excess in $\Delta(I-K)$.  We have 15 H$\alpha$ emitting periodic variables that are marked in Fig~\ref{fig:per_IRexcess}.  These  H$\alpha$ emitting sources are selected from IPHAS photometry (see sec. \ref{sec:h_alpha}). NIR excess indicates the presence of near-photospheric circumstellar dust, while H$\alpha$ emission is a good indicator of accretion in T-Tauri stars. Out of 16 NIR excess sources, 10 objects exhibit H$\alpha$ emission.  We obtained disk fraction of about 41\% among periodic variables from the NIR excess analysis. Since we have longer wavelengths Spitzer data in IRAC bands 3.6 and 4.5 $\micron$ for NGC~2282 as mentioned in sec. \ref{data_sets} (see also Paper I), these could be used for probing better disk indicator than  NIR wavelengths \citep[][]{2000AJ....120.3162L,2010A&A...515A..13R,2017ApJ...836...98J}. However, nebular PAH emission on the central part of the cluster might be problematic sometimes at IRAC bands \citep[][]{2009ApJS..184...18G}. Among 41 periodic variables, 33 have MIR data.  In the lower panel of Fig.\ref{fig:per_IRexcess}, we plotted the period versus $Spitzer$-IRAC (see. section 2) color [3.6] -[4.5] for all 33 variables. The population of variables close to [3.6] -[4.5]$\approx$ 0 indicates diskless bare photospheric color \citep[][]{2006ApJ...651..502P}. The colors [3.6] -[4.5] $>$ 0.7 and $>$0.15 indicate Class I and Class II sources respectively \citep[][]{2008ApJ...674..336G,2009ApJS..184...18G}. We adopt conservatively the color [3.6] -[4.5] $>$ 0.25   criterion for the disked stars as shown in Fig.\ref{fig:per_IRexcess} (lower panel). We obtained relatively higher disk fraction of $\sim$51\% based on MIR data. This is comparable to the disk fraction estimation in Paper I ($\sim$58\%) for all the PMS sources within  NGC 2282. 

\subsubsection{Rotation/variability amplitude to disk connection}

\begin{figure}
\includegraphics[width=7.5 cm,height=8.5 cm, angle=270]{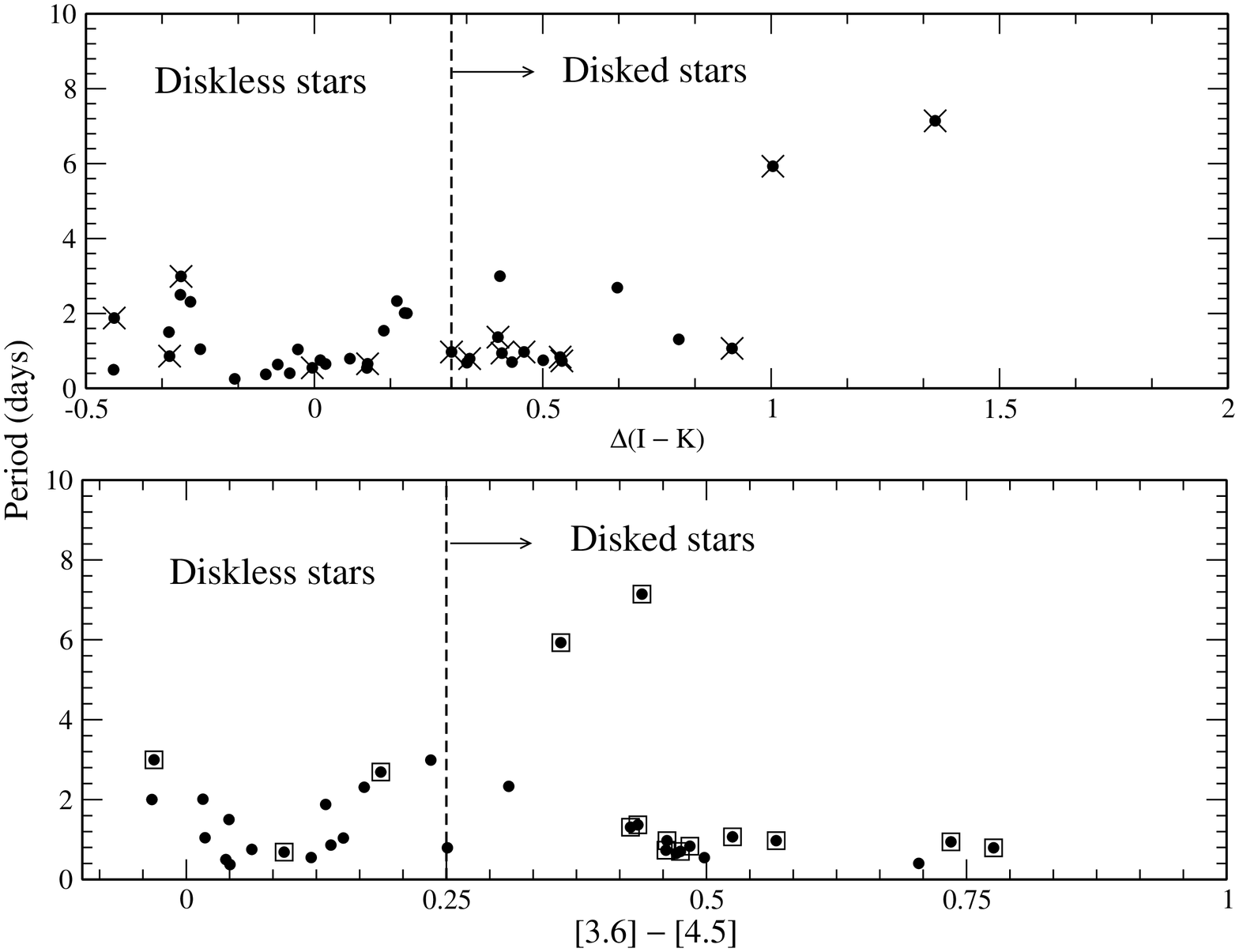}

\caption{Periods are plotted  as a function of disk indicator NIR excess $\Delta(I - K)$ (upper panel) and MIR excess [3.6]$-$[4.5] (lower panel) for periodic variables. The vertical solid lines of $\Delta(I-K)$ = 0.3 and [3.6]$-[$4.5]=0.25 are adopted as the boundary of diskless and disked stars. In upper panel, the H$\alpha$ emitting sources are marked by the cross symbols. In lower panel, NIR excess sources are marked by the   squares. No conclusive trends are seen  in the plots for disk-locking mechanism.}
  \label{fig:per_IRexcess}
\end{figure}

To understand rotation-disk connection in our periodic variables, the IR excesses versus periods are plotted in Fig.\ref{fig:per_IRexcess}.  The disk-locking system should have relatively longer periods as compared to that of diskless stars as studied before in the literature. In Fig.\ref{fig:per_IRexcess}, the few disked stars have a larger range of periods whereas the few slow rotators also have disks around them. This finding could be related to the age of NGC~2282, as it is relatively old (2$-$3 Myrs), and such disk-locking systems may be efficient at the young age of $\sim$ 1 Myr or less as suggested by several authors and already described in earlier section 3.3. However, we are unable to draw any conclusion about disk-locking theory due to our statistically small number of samples.

\begin{figure}
\includegraphics[width=7.5 cm,height=8.5 cm, angle=270]{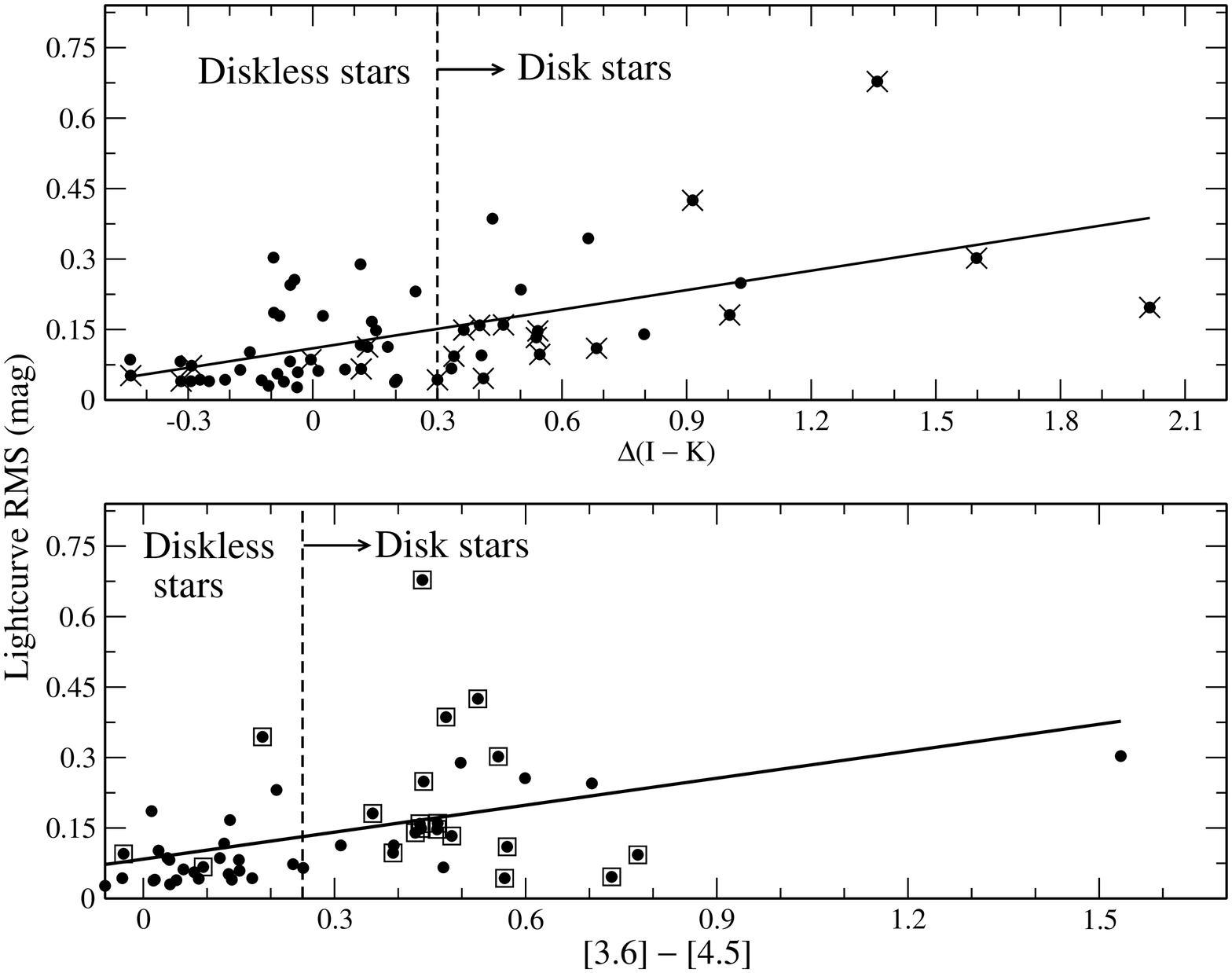}

\caption{The light curves RMS (variability amplitude) are plotted  as a function of disk indicator NIR excess $\Delta(I - K)$ (upper panel) and MIR excess [3.6]$-$[4.5] (lower panel) for periodic and aperiodic variables. The vertical dotted lines of $\Delta(I-K)$ = 0.3 and [3.6]$-$[4.5]=0.25 are adopted as the boundary of diskless and disked stars. In the upper panel, the H$\alpha$ emitting sources are marked by the cross symbols. In the lower panel, NIR excess sources are marked by the open squares. The dashed line obeys a clear increasing trend of RMS with infrared excess.}
  \label{fig:RMS_IKspitzer}
\end{figure}

 We  also analyzed whether there is any correlation between variability amplitude of the periodic/aperiodic variables and the IR excess using $\Delta(I-K)$ and [3.6]-[4.5] as stated in the above section. We have total 62 variables, and 51 have both [3.6] and [4.5] measurements.  In Fig.~\ref{fig:RMS_IKspitzer}, we plotted the variability amplitude from the light curve RMS versus NIR excess $\Delta(I-K)$ (upper panel) and MIR excess [3.6]-[4.5] (lower panel). A trend of increasing variability amplitude with IR excess is seen in Fig.\ref{fig:RMS_IKspitzer}.  
Such results suggest that the objects with thick circumstellar disk tend to have relatively larger light-curve amplitudes.  Thus, the presence of thick disk around stars preferably display variable phenomena due to their accretion activity.

\subsection{Period-mass correlation}

\begin{figure}
\includegraphics[width=8.0 cm,height=6.0 cm]{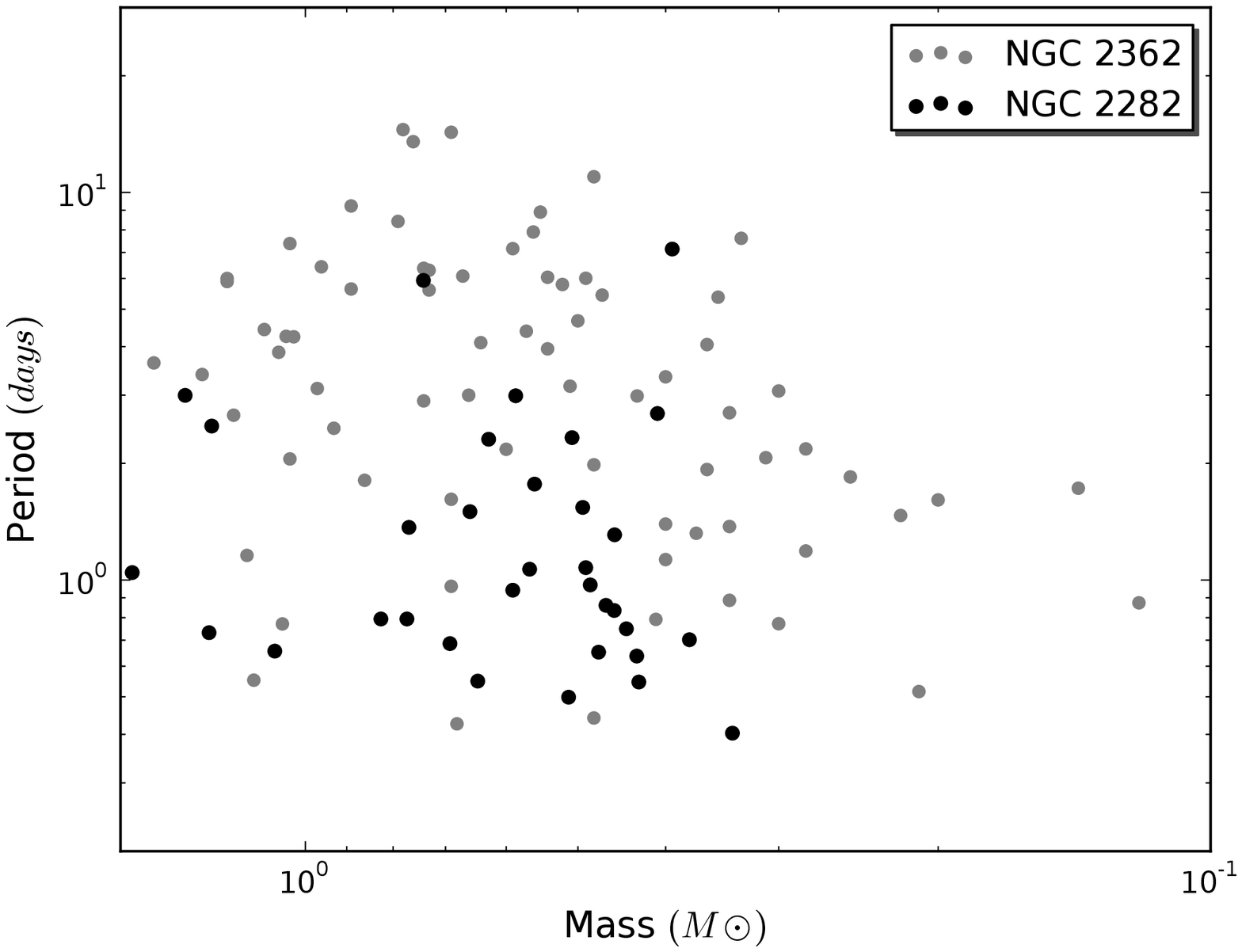}

\caption{Mass versus period is plotted for periodic variables in the mass range $\sim$0.25$-$1.6 M$_\odot$. From our detection limit, periodic variables are identified up to the low mass limit of $\geq$ 0.25 M$_\odot$, and a systematic trend for fast rotators at the low mass-end might be real feature as seen in other a few regions \citep[see][]{2008MNRAS.384..675I}}
  \label{fig:per_mass}
\end{figure}

To investigate any correlation between the rotation periods and the masses, we have plotted the rotation periods versus stellar masses for periodic variables in NGC~2282 in Fig.\ref{fig:per_mass}. The masses are estimated from I-band magnitudes and the mean cluster extinction (A$_v$=1.65 mag), based on the 3 Myr theoretical isochrones of \citet[][]{1998A&A...337..403B}. The estimated masses range from 0.1 M$\odot$ to 2.5 M$\odot$.  The mass estimation based on the isochrones is quite uncertain at these young ages. Furthermore, the lower mass limit of 0.1 M$\odot$  lies near to I$\sim$ 20.5 mag, where the light curves do not have sufficient precision to detect any variability due to poor SNR at such faint end.  For comparison, we plotted the period distribution with masses in NGC 2362 ($\sim$5 Myr), taken from \citet[][]{2008MNRAS.384..675I}. From Fig.\ref{fig:per_mass}, it is apparent that the rotation period follows a flat to sloping mass dependence of the rotation periods in NGC 2282 ($\sim$3 Myr). However, our periodic variables detection is not sensitive enough at the lower mass close to 0.3 M$\odot$, and hence the result has to be taken with caution. Furthermore, we suspect that the mass determinations might be unreliable due to spatial variable extinction in the cluster area. 
Such variable extinction introduces horizontal dispersion in the mass versus period plot. Lack of spectroscopy on such individual low-mass objects, the proper value of extinction may be unreliable from photometry only, and we prefer to use the mean extinction to estimate masses here.  It is worth to point out that by adopting variable extinction values in the individual cluster members would cause the changes in the stellar masses  25 $-$ 40\%  of our estimated values.

Our result agrees with the correlation of gradual evolutionary sequence;  no mass dependence with the rotation periods at the younger cluster to the mass dependent rotation rates at the older cluster \citep[e.g.][and references therein]{2008MNRAS.384..675I,2016ApJ...833..122C}. Fig.~12 of \citet[][]{2008MNRAS.384..675I} shows no-mass dependency relation to the rotation periods in the ONC ($\sim$1 Myr). While a weak-mass dependency (fast rotation at the low-mass stars compared to that of high-mass stars) relation is seen in NGC 2264 ($\sim$2 Myr) and 2362($\sim$5 Myr), and a strong-mass dependency in more older clusters NGC 2547 ($\sim$40 Myr) and NGC 2516 ($\sim$150 Myr).

\section{Summary and Conclusions}

In this paper, we present deep $I$-band ($\sim$ 20.5 mag) long-term photometric monitoring studies of the stars towards a young (2$-$5 Myr) cluster NGC 2282 in the Monoceros constellation to understand the variability characteristics towards the low-mass end of PMS stars. Our main results are summarized as follows:
 \begin{enumerate}

 \item From the differential photometry light curves of 1627 stars, we have identified 62 new photometric variable stars.  Out of 62 variables, 41 are periodic variables and  yield the rotation periods from 0.2 to 7 days.

 \item Based on the  H$\alpha$ emission activities, J$-$H/H$-$K CC diagram, IR excess, and paper I, we found the majority of the variable stars are members of the cluster. 
 
\item The period distribution shows a median period of $\sim$1 day as seen in other young clusters (e.g., NGC~2264, ONC,  etc.). It displays a unimodal distribution like others;   but younger clusters like the ONC have bimodal distribution with slow rotators peaking at $\sim$ 6$-$8 days.

\item To understand the disk-locking hypothesis in young PMS though rotation-disk connection, we derived NIR excess from $\Delta$(I$-$K) and MIR excess from $Spitzer$ [3.6]$-$[4.5] $\mu$m data. No conclusive  correlation of slow rotation  with the presence of disks around stars and fast rotation for diskless stars is seen from the periodic variables in NGC~2282. 

\item Furthermore,  to investigate the variability-disk connection, we studied NIR excess, and MIR excess from $Spitzer$ data with the variability amplitude (RMS) derived from the light curves. A clear increasing trend of variability amplitude with IR excess is seen for all the periodic/aperiodic variables.

\item Disk fraction among the periodic variables are studied using NIR and MIR excess as discussed, we estimated the fraction of $\sim$51\% for all the periodic/aperiodic variables, while $\sim$41\% for periodic variables. This estimation is comparable to the disk fraction ($\sim$58\%) obtained for the entire PMS population in Paper I.

\item To understand fast rotation in the lower mass end of the PMS stars, the period-mass correlation in the mass range 0.3$-$1.6 M$_\odot$ is studied. It is apparent from our analysis that there is an evidence of relatively fast rotation with decreasing masses (sloping relation) as seen in the literature \citep[][]{2008MNRAS.384..675I,2016ApJ...833..122C}. However, our periodic variables selection is not sensitive enough at the lower mass beyond 0.3 M$_\odot$, and hence the results need to be taken with caution.  
 \end{enumerate}

\section*{Acknowledgments}

The authors are very much thankful to the anonymous referee for his/her critical and valuable comments, which help us to improve the paper. This research work is supported by the Satyendra Nath Bose National Centre for Basic Sciences under the Department of Science and Technology, Govt. of India. This paper is based on data obtained as part of the UKIRT Infrared Deep Sky Survey. We also use data product of Two-micron All-Sky Survey (2MASS), which is a joint project of the University of Massachusetts and Infrared Processing and Analysis Center/California Institute of Technology, funded by National Aeronautics and Space Administration [NASA] and the National Science Foundation). This publication also made use of observations with the {\it Spitzer Space Telescope} (operated by the Jet Propulsion Laboratory, California Institute of Technology, under contract with NASA). The authors are grateful to the HTAC members and staff of HCT, operated by Indian Institute of Astrophysics (Bangalore); JTAC members and staff of 1.30m DFOT, operated by Aryabhatta Research Institute of Observational Sciences ( Nainital).

\bibliographystyle{mnras}
\bibliography{ngc2282}

\begin{table*}
\renewcommand{\tabcolsep}{2.5pt}
\caption{Catalog of the identified variable stars. The complete table is available in the electronic version.}
\begin{threeparttable}

\label{tab:cat_var}
\begin{tabular}{cccccccccccccc}
\hline \multicolumn{1}{c}{ID} & \multicolumn{1}{c}{$\alpha_{2000}$} & \multicolumn{1}{c}{$\delta_{2000}$} & \multicolumn{1}{c}{$B$} & \multicolumn{1}{c}{$V$}& \multicolumn{1}{c}{$R$} & \multicolumn{1}{c}{$I$} & \multicolumn{1}{c}{$J$} & \multicolumn{1}{c}{$H$} & \multicolumn{1}{c}{$K$} & \multicolumn{1}{c}{3.6 $\mu$m} & \multicolumn{1}{c}{4.5 $\mu$m}& \multicolumn{1}{c}{W1} & \multicolumn{1}{c}{W2}\\ 

\multicolumn{1}{c}{} & \multicolumn{1}{c}{(deg)} & \multicolumn{1}{c}{(deg)} & \multicolumn{1}{c}{(mag)} & \multicolumn{1}{c}{(mag)}& \multicolumn{1}{c}{(mag)} & \multicolumn{1}{c}{(mag)} & \multicolumn{1}{c}{(mag)} & \multicolumn{1}{c}{(mag)} & \multicolumn{1}{c}{(mag)} & \multicolumn{1}{c}{(mag)} & \multicolumn{1}{c}{(mag)}& \multicolumn{1}{c}{(mag)} & \multicolumn{1}{c}{(mag)}\\ \hline

    8 &  101.735121  &  1.277934  &  16.971 & 15.934 & 14.935 & 14.189 & 12.061 & 11.212 & 10.572 &   9.616 &   9.049 &   4.189 &   1.079 \\
  &    &    &  $\pm$ 0.012 & $\pm$ 0.013 & $\pm$ 0.007 & $\pm$ 0.003 & $\pm$ 0.023 & $\pm$ 0.020 & $\pm$ 0.023 & $\pm$ 0.003 & $\pm$ 0.002 & $\pm$ 0.015 & $\pm$ 0.021  \\
   16 &  101.697027  &  1.251858  &  17.546 & 16.037 & 15.260 & 14.314 & 12.930 & 12.423 & 12.075 &  12.083 &  99.999 &  10.109 &   7.073 \\
  &    &    &  $\pm$ 0.049 & $\pm$ 0.043 & $\pm$ 0.039 & $\pm$ 0.046 & $\pm$ 0.026 & $\pm$ 0.034 & $\pm$ 0.034 & $\pm$ 0.004 & $\pm$99.999 & $\pm$ 0.092 & $\pm$ 0.160  \\
   38 &  101.681203  &  1.275360  &  15.257 & 14.608 & 14.147 & 13.876 & 13.306 & 12.956 & 12.952 &  12.842 &  12.826 &   9.541 &   7.619 \\
  &    &    &  $\pm$ 0.007 & $\pm$ 0.008 & $\pm$ 0.009 & $\pm$ 0.005 & $\pm$ 0.030 & $\pm$0.001 & $\pm$ 0.044 & $\pm$ 0.008 & $\pm$ 0.009 & $\pm$ 0.093 & $\pm$ 0.203  \\
   52 &  101.679695  &  1.290380  &  19.691 & 17.640 & 16.207 & 14.898 & 12.795 & 11.770 & 11.406 &  11.121 &  11.154 &  11.955 &   8.283 \\
  &    &    &  $\pm$ 0.020 & $\pm$ 0.004 & $\pm$ 0.009 & $\pm$ 0.006 & $\pm$ 0.020 & $\pm$ 0.026 & $\pm$ 0.023 & $\pm$ 0.003 & $\pm$ 0.004 & $\pm$99.999 & $\pm$99.999  \\
   53 &  101.715712  &  1.307466  &  16.823 & 15.588 & 14.729 & 13.999 & 12.775 & 12.126 & 11.936 &  11.688 &  11.646 &   6.786 &   3.643 \\
  &    &    &  $\pm$ 0.008 & $\pm$ 0.004 & $\pm$ 0.007 & $\pm$ 0.003 & $\pm$ 0.026 & $\pm$ 0.028 & $\pm$ 0.030 & $\pm$ 0.004 & $\pm$ 0.005 & $\pm$99.999 & $\pm$ 0.119  \\
   54 &  101.723024  &  1.310400  &  18.770 & 17.447 & 16.363 & 17.674 & 13.677 & 12.644 & 11.958 &  10.761 &  10.026 &  99.999 &  99.999 \\
  &    &    &  $\pm$ 0.013 & $\pm$ 0.005 & $\pm$ 0.009 & $\pm$ 0.009 & $\pm$ 0.026 & $\pm$ 0.026 & $\pm$ 0.027 & $\pm$ 0.003 & $\pm$ 0.002 & $\pm$99.999 & $\pm$99.999  \\
   55 &  101.714484  &  1.319228  &  99.999 & 16.760 & 16.161 & 17.549 & 13.439 & 12.624 & 12.409 &  12.229 &  12.058 &  99.999 &  99.999 \\
  &    &    &  $\pm$99.999 & $\pm$ 0.072 & $\pm$ 0.008 & $\pm$ 0.020 & $\pm$ 0.036 & $\pm$ 0.027 & $\pm$ 0.033 & $\pm$ 0.004 & $\pm$ 0.006 & $\pm$99.999 & $\pm$99.999  \\
   58 &  101.718528  &  1.316127  &  16.417 & 15.134 & 14.292 & 13.575 & 12.280 & 11.593 & 11.379 &  11.219 &  11.139 &  99.999 &  99.999 \\
  &    &    &  $\pm$ 0.014 & $\pm$ 0.011 & $\pm$ 0.012 & $\pm$ 0.004 & $\pm$ 0.029 & $\pm$ 0.033 & $\pm$ 0.033 & $\pm$ 0.003 & $\pm$ 0.004 & $\pm$99.999 & $\pm$99.999  \\
 \hline\end{tabular}
\begin{tablenotes}\footnotesize
 \item [] N.B. The value `99.999' represent the absence of the particular magnitude.
\end{tablenotes}
\end{threeparttable}
\end{table*}

\clearpage


\begin{table*}
\renewcommand{\tabcolsep}{2.5pt}
\caption{Result of our analysis : periods, membership characteristics, IR excess, masses  of the identified variable stars.}
\begin{threeparttable}

\label{tab:var_phy}
\begin{tabular}{ccccccccccc}
\hline \multicolumn{1}{c}{ID} & \multicolumn{1}{c}{I\tnote{*}} & \multicolumn{1}{c}{I$_{err}$} & \multicolumn{1}{c}{RMS} & \multicolumn{1}{c}{H$\alpha$\tnote{**}} &\multicolumn{1}{c}{Period} &\multicolumn{1}{c}{PMS from Paper I}  &\multicolumn{1}{c}{$\Delta$(I$-$K)}  &\multicolumn{1}{c}{[3.6]$-$[4.5]}  &\multicolumn{1}{c}{mass\tnote{***}}\\ 

\multicolumn{1}{c}{} & \multicolumn{1}{c}{(mag)} & \multicolumn{1}{c}{(mag)} & \multicolumn{1}{c}{(mag)}& \multicolumn{1}{c}{(emission Y/N)}& \multicolumn{1}{c}{(days)}& \multicolumn{1}{c}{(Yes/No)} & \multicolumn{1}{c}{(mag)} & \multicolumn{1}{c}{(mag)}&  \multicolumn{1}{c}{(M$_\odot$)}\\ \hline
8	&	14.189	&	0.003	&	0.043	&	Y	&	0.972	&	Yes	&	0.300	&	0.567	&	2.43	\\
16	&	14.314	&	0.046	&	0.064	&	N	&	0.254	&	No	&	-0.174	&	$...$	&	2.33	\\
38	&	13.876	&	0.005	&	0.038	&	N	&	2.011	&	No	&	0.197	&	0.016	&	2.69	\\
52	&	14.898	&	0.006	&	0.043	&	N	&	2.003	&	No	&	0.202	&	-0.033	&	1.90	\\
53	&	13.999	&	0.003	&	0.03	&	N	&	0.373	&	No	&	-0.106	&	0.042	&	2.58	\\
54	&	17.674	&	0.009	&	0.046	&	Y	&	0.942	&	Yes	&	0.410	&	0.735	&	0.59	\\
55	&	17.549	&	0.02	&	0.043	&	N	&	2.309	&	No	&	-0.271	&	0.171	&	0.63	\\
58	&	13.575	&	0.004	&	0.056	&	N	&	$...$	&	No	&	-0.085	&	0.08	&	2.96	\\
60	&	15.274	&	0.005	&	0.059	&	N	&	1.039	&	No	&	-0.036	&	0.151	&	1.66	\\
62	&	14.979	&	0.004	&	0.062	&	N	&	0.754	&	No	&	0.013	&	0.063	&	1.84	\\
115	&	15.27	&	0.008	&	0.052	&	Y	&	1.879	&	Yes	&	-0.438	&	0.134	&	1.66	\\
226	&	18.026	&	0.058	&	0.148	&	N	&	1.540	&	No	&	0.152  &	$...$	&	0.49	\\
279	&	15.965	&	0.009	&	0.04	&	N	&	2.497	&	No	&	-0.293	&	$...$	&	1.27	\\
321	&	16.448	&	0.008	&	0.249	&	N	&	$...$	&	Yes	&	1.030	&	0.44	&	1.04\\
363	&	18.182	&	0.025	&	0.14	&	N	&	1.309	&	Yes	&	0.797	&	0.427	&	0.45	\\
364	&	18.299	&	0.044	&	0.289	&	N	&	0.546	&	Yes	&	0.114	&	0.498	&	0.43\\
366	&	17.17	&	0.015	&	0.082	&	N	&	$...$	&	No	&	-0.054	&	0.15	& 0.75	\\
382	&	16.968	&	0.012	&	0.065	&	N	&	0.794	&	Yes	&	0.077	&	0.251	&	0.82	\\
390	&	17.202	&	0.013	&	0.181	&	Y	&	5.930	&   Yes	&  1.003 	&	0.36	&	0.74	\\
480	&	15.747	&	0.007	&	0.042	&	N	&	$...$	&	No	&	-0.122	&	0.087	&	1.38	\\
506	&	17.956	&	0.028	&	0.086	&	N	&	0.499	&	No	&	-0.439	&	0.038	&	0.51	\\
521	&	17.451	&	0.02	&	0.082	&	N	&	1.502	&	No	&	-0.318	&	0.041	&	0.65	\\
566	&	17.345	&	0.021	&	0.067	&	N	&	0.686	&	No	&	0.333	&	0.094	&	0.69	\\
571	&	16.35	&	0.01	&	0.097	&	Y	&	$...$	&	Yes	&	0.546	&	0.392	&	1.08\\
575	&	17.76	&	0.021	&	0.425	&	Y	&	1.066	&	Yes	&	0.914	&	0.525	&	0.56	\\
578	&	19.061	&	0.026	&	0.113	&	Y	&	$...$	&	Yes	&	0.132	&	0.393	&	0.28\\
581	&	18.14	&	0.016	&	0.04	&	Y	&	0.861	&	Yes	&	-0.317	&	0.139	&	0.46	\\
603	&	16.54	&	0.009	&	0.11	&	Y	&	$...$	&	Yes	&	0.683	&	0.571	&	0.99\\
625	&	18.063	&	0.028	&	0.16	&	Y	&	0.972	&	Yes	&	0.459	&	0.462	&	0.48	\\
629	&	17.367	&	0.018	&	0.149	&	Y	&	$...$	&	Yes	&	0.363	&	0.436	&	0.68	\\
636	&	16.062	&	0.009	&	0.039	&	N	&	$...$	&	No	&	-0.069	&	0.052	&	1.22	\\
642	&	17.178	&	0.013	&	0.117	&	N	&	$...$	&	No	&	0.115	&	0.127	&	0.75	\\
644	&	17.396	&	0.018	&	0.302	&	Y	&	$...$	&	Yes	&	1.597	&	0.557	&	0.67	\\
652	&	16.584	&	0.01	&	0.167	&	N	&	$...$	&	Yes	&	0.142	&	0.136	&	0.98	\\
654	&	18.388	&	0.025	&	0.344	&	N	&	2.690	&	Yes	&	0.663	&	0.187	&	0.41	\\
677	&	17.123	&	0.015	&	0.159	&	Y	&	1.368	&	Yes	&	0.401	&	0.434	&	0.77	\\
841	&	14.575	&	0.004	&	0.027	&	N	&	$...$	&	No	&	-0.037	&	-0.06	&	2.13	\\
976	&	15.449	&	0.007	&	0.04	&	N	&	1.046	&	No	&	-0.249	&	0.018	&	1.55	\\
998	&	15.796	&	0.02	&	0.095	&	N	&	2.996	&	No	&	0.406	&	-0.031	&	1.35	\\
999	&	18.18	&	0.04	&	0.133	&	Y	&	0.835	&	Yes	&	0.538	&	0.484	&	0.45	\\
1060	&	16.859	&	0.013	&	0.231	&	N	&	$...$	&	Yes	&	0.247	&	0.209	&	0.86	\\
1144	&	16.113	&	0.037	&	0.197	&	Y	&	$...$	&	Yes	&	2.014	&	$...$	&	1.19	\\
1194	&	18.104	&	0.062	&	0.179	&	N	&	0.652	&	No	&	0.024	&	$...$	&	0.47	\\
1244	&	18.289	&	0.047	&	0.179	&	N	&	0.637	&	No	&	-0.080	&	$...$	&	0.43	\\
1366	&	17.689	&	0.045	&	0.073	&	Y	&	2.989	&	Yes	&	-0.292	&	0.235	&	0.58	\\
1876	&	18.539	&	0.048	&	0.386	&	N	&	0.702	&	No	&	0.432	&	0.475	&	0.37	\\
2286	&	17.492	&	0.032	&	0.086	&	Y	&	0.549	&	Yes	&	-0.004	&	0.12	&	0.64	\\
2301	&	17.914	&	0.04	&	0.256	&	N	&	$...$	&	Yes	&	-0.043	&	0.599	&	0.52 \\
2303	&	15.948	&	0.047	&	0.147	&	Y	&	0.732	&	Yes	&	0.541	&	0.461	&	1.27	\\
2319	&	18.739	&	0.045	&	0.245	&	N	&	0.403	&	No	&	-0.053	&	0.704	&	0.34	\\
2329	&	17.112	&	0.028	&	0.093	&	Y	&	0.794	&	Yes	&	0.339	&	0.776	&	0.77	\\
2337	&	16.353	&	0.007	&	0.066	&	Y	&	0.656	&	Yes	&	0.116	&	0.471	&	1.08\\
2363	&	17.762	&	0.01	&	0.186	&	N	&	$...$	&	No	&	-0.093	&	0.013	&	0.56	\\
2424	&	19.398	&	0.143	&	0.303	&	N	&	$...$	&	No	&	-0.094	&	1.534	&	0.23	\\
2444	&	18.458	&	0.041	&	0.678	&	Y	&	7.143	&	Yes	&	1.359	&	0.438	&	0.39	\\
2692	&	16.681	&	0.019	&	0.102	&	N	&	$...$	&	No	&	-0.151	&	0.024	&	0.94	\\
3045	&	17.973	&	0.033	&	0.113	&	N	&	2.331	&	Yes	&	0.180	&	0.31	&	0.51	\\
3954	&	18.239	&	0.088	&	0.235	&	N	&	0.749	&	No	&	0.501	&	$...$	&	0.44	\\
6001	&	15.348	&	0.008	&	0.043	&	N	&	$...$	&	No	&	-0.210	&	$...$	&	1.61 \\
6002	&	17.785	&	0.026	&	0.111	&	N	&	1.769	&	No	&	$...$	&	$...$	&	0.56	\\
6003	&	18.041	&	0.067	&	0.222	&	N	&	1.077	&	No	&	$...$	&	$...$	&	0.49	\\
6004	&	17.744	&	0.046	&	$...$	&	N	&	$...$	&	No	&	$...$	&	$...$	&	0.57	\\
\hline
\end{tabular}
\begin{tablenotes}\footnotesize
 \item [*] Mean $I$ magnitude from the light curve.\\
 \item [**] From IPHAS photometry. See text for details.\\
\item [***] From I magnitude, mean extinction and theoretical isochrones of 3 Myr from \citet[]{1998A&A...337..403B} \\
\end{tablenotes}
\end{threeparttable}
\end{table*} 





\end{document}